\documentclass{article}

\usepackage{csquotes} % must be before biblatex
\usepackage[
  backend=biber,
  style=apa,
  maxcitenames=99,   % show all authors in citations
  mincitenames=1,    % never collapse to "et al."
  maxbibnames=99,    % show all authors in bibliography
  giveninits=true     % initials instead of full given names (optional)
]{biblatex}
\DeclareLanguageMapping{american}{american-apa}
\usepackage[a4paper, total={6in, 9in}]{geometry}
\usepackage{setspace} 
\usepackage[graphicx]{realboxes}
\usepackage{longtable}
\usepackage{colortbl}
\usepackage{rotating}
\usepackage{array}
\usepackage{mathtools}
\usepackage{multicol}
\usepackage{multirow}
\usepackage{comment}
\usepackage{caption}
\usepackage{subcaption}
\usepackage{bbm}
\usepackage{tikz}
\usepackage{amsmath,amssymb}

\usepackage{tabularx}
\newcolumntype{Z}{>{\raggedright\arraybackslash}X}
\newcolumntype{Y}{>{\centering\arraybackslash}X}

\usepackage[nomarkers, nolists, figuresfirst]{endfloat}

\usepackage[colorlinks]{hyperref}
\usepackage[nameinlink, noabbrev]{cleveref}
\hypersetup{colorlinks = true, linkcolor = blue, citecolor = blue}

\title{Beyond Test Scores: \\ How Academic Rank Shapes Long-Term Outcomes%
  \thanks{\normalsize{We are grateful to Massimiliano Bratti, Margherita Fort,
  Rigissa Megalokonomou, Birgitta Rabe, Daniela Sonedda, and especially Emma
  Duchini and Ingo Isphording for the precious feedback and suggestions provided
  on this study. We also appreciate the constructive comments from participants
  at the 2024 Workshop in Economics of Education and Policy (WEEP), the 2024
  EEA-ESEM Conference, and seminar attendees at ISER, Curtin University, and
  Monash University. We thank Heather Clark (University of Aberdeen) and the team
  at the Grampian Data Safe Haven (DaSH) for their assistance with data access
  and management. The findings and opinions expressed in this paper are solely our
  own and do not necessarily reflect those of DaSH or the University of Aberdeen.
  We acknowledge funding from the ESRC Research Centre on Micro-Social Change
  (MiSoC), award number ES/S012486/1. This paper also benefited from the use of
  OpenAI's ChatGPT for brainstorming, drafting, and writing assistance. All errors
  are our own. \\[4pt]
  \small $^{\dagger}$Institute for Social and Economic Research, University of
  Essex. Email: \href{mailto:edelbono@essex.ac.uk}{edelbono@essex.ac.uk} \\[-2pt]
  \small $^{\ddagger}$Institute for Social and Economic Research, University of
  Essex. Email: \href{mailto:ajholf@essex.ac.uk}{ajholf@essex.ac.uk} \\[-2pt]
  \small $^{\S}$Department of Economics, Monash University. Email:
  \href{mailto:tommaso.sartori@monash.edu}{tommaso.sartori@monash.edu}}}}

\author{Emilia Del Bono$^{\dagger}$ \and Angus Holford$^{\ddagger}$ \and
Tommaso Sartori$^{\S}$}

\begin{document}

\maketitle
\pagenumbering{gobble}

\begin{abstract}

We study the effects of academic rank using data on the entire population of children enrolled in Aberdeen's (Scotland) primary schools in 1962. Exploiting quasi-random variation in peer group composition, we estimate the causal impact of rank on academic performance, non-cognitive development, parental investment, and long-term outcomes. A higher rank improves achievement on the high-stakes 11-plus exam and enhances internalizing skills---traits linked to self-concept and confidence---suggesting that rank effects operate primarily through students’ self-perception. Using a follow-up survey conducted 40 years later, we find that rank raises educational attainment, particularly for girls, but long-term income gains emerge only among boys. The gender gap in long-run effects likely reflects historical constraints on women’s access to higher education and skilled employment during this period.

\end{abstract}


%%%%%%%%%%%%%%%%%%%%
% Acknowledgements %
%%%%%%%%%%%%%%%%%%%%

%\footnote{\normalsize{We are grateful to Massimiliano Bratti, Margherita Fort, Rigissa Megalokonomou, Birgitta Rabe, Daniela Sonedda, and especially Emma Duchini and Ingo Isphording for the precious feedback and suggestions provided on this study. We also appreciate the constructive feedback from participants at the 2024 Workshop in Economics of Education and Policy (WEEP), the 2024 EEA-ESEM Conference, and seminar attendees at ISER, Curtin University, and Monash University. We thank Heather Clark (University of Aberdeen) and the team at the Grampian Data Safe Haven (DaSH) for their assistance with data access and management. The findings and opinions expressed in this paper are solely our own and do not necessarily reflect those of DaSH or the University of Aberdeen. We acknowledge funding from the ESRC Research Centre on Micro-Social Change (MiSoC), award number ES/S012486/1. Finally, this paper also benefited from the use of OpenAI's ChatGPT for brainstorming, drafting, and writing assistance. All remaining errors are our own.}}}

\clearpage \pagenumbering{arabic}

%%%%%%%%%%%%%%%%
% Introduction %
%%%%%%%%%%%%%%%%

\section{Introduction} \label{sec_1}

\vspace{1mm}
Decades of research on the impact of peers on academic outcomes have generally identified positive but moderate effects, despite extensive debate over the appropriate methodological framework.\footnote{An extensive debate on the appropriate functional form for estimating peer effects has highlighted the limitations of the classic linear-in-means model, which may overlook important within-group heterogeneity, ignore selection effects, and fail to capture the dynamic nature of peer influence. While alternative approaches often find more nuanced and sometimes stronger effects, these differences are typically modest. For a comprehensive review, see \textcite{sacerdote2011}.} While these findings suggest that direct peer effects may be limited, a parallel literature on students' academic rank has consistently documented much larger impacts, revealing a different channel through which peer groups shape individual outcomes. Building on the ``big fish in a little pond'' theory \parencite{marsh1984}, this body of work argues that students' relative standing within their peer group can have substantial and lasting consequences for their future trajectories.
 
\vspace{1mm}
More specifically, researchers have found that higher academic rank positively influences future academic performance, educational attainment, career choice \parencite{elsner2017, elsner2018, murphy2020, elsner2021, pagani2021, megalokonomou2024, carneiro2023, denning2023}, major choice \parencite{delaney2021, goulas2022}, and future earnings \parencite{denning2023}. We contribute to this growing literature by extending the range of outcomes considered, the measurement approaches employed, and the mechanisms explored — delving more deeply into the non-cognitive channels through which rank operates, examining its relationship with parental investment, and tracing its long-term impacts across a broad set of educational, socioeconomic, and well-being dimensions.

\vspace{1mm}
We study the effects of academic rank using a unique dataset covering the full population of children born between October 1950 and October 1955 in Aberdeen, Scotland, and enrolled in the city’s primary schools. The dataset includes nearly 10,000 children and draws on three rich sources of information. First, a comprehensive survey conducted between 1962 and 1964 gathered data on children's academic progression, cognitive performance, and anthropometric indicators through school medical exams and hospital birth records. Non-cognitive skills were measured using the Rutter Questionnaire \parencite{rutter1967}, a widely validated behavioral assessment completed by teachers. Second, a randomly selected 20\% subsample of children was surveyed at home, with parents providing detailed information on family structure, social and health behaviors, parental involvement, and time use. Finally, a follow-up survey conducted in 2001 (with the original participants aged between 46 and 51 years old) achieved a response rate of nearly 60\%, offering detailed information on adult outcomes including education, occupation, income, health, and subjective well-being. The breadth and depth of this dataset allow us to trace the effects of academic rank from childhood well into midlife.

\vspace{1mm}
We construct academic rank based on students’ performance on a standardized cognitive test administered around age 9, and define each student’s peer group at the school-cohort level — typically involving 30–35 children, small enough to ensure meaningful peer interaction. Our identification strategy exploits quasi-random variation in rank within these groups: we show that group composition is conditionally random, with no evidence of systematic sorting based on observed characteristics. This allows us to interpret within-group rank as plausibly exogenous to underlying ability, parental background, or non-cognitive traits. Our analysis then examines how this measure of rank influences a range of outcomes, from academic performance and behavioral skills in primary school to parental investment and long-term educational and economic trajectories.

\vspace{1mm}
We find that academic rank has a substantial and robust impact on students’ academic performance, as measured by the results of the 11-plus examination — a high-stakes standardized test taken at the end of primary school, which determined access to grammar schools, the selective secondary track in the UK system. A rise of four positions in a student’s rank within their school-cohort group — equivalent to a 10\% improvement in relative standing given the average group size — raises 11-plus test scores by roughly 6\% of a standard deviation, holding cognitive skills, individual, and peer characteristics constant. Interpreting rank effects linearly is supported by the data: we show that the estimated relationship is approximately linear across the full rank distribution, consistent with findings in \textcite{murphy2020} and \textcite{elsner2021}, and suggestive of a self-preserving dynamic where early rank boosts performance, leaving the rank unchanged. Moreover, the effect is stronger for girls, who not only begin with higher average test scores but also respond more sharply to their relative group position — an important pattern that echoes the gender differences in sensitivity to feedback observed in other peer-related settings \parencite{lavy2019, buser2023}.

\vspace{1mm}
To study the effect of academic rank on non-cognitive development, we derive two standardized measures — externalizing and internalizing skills — using factor analysis on teacher responses to the Rutter Questionnaire \parencite{rutter1967}. This method allows us to reduce dimensionality and recover latent behavioral traits from a broad set of teacher-reported items, following a well-established approach in developmental psychology \parencite[e.g.,][]{boyle1985, mcgee1985}, and increasingly adopted in economics \parencite[e.g.,][]{attanasio2020}. Externalizing skills capture outward-oriented behaviors such as restlessness, aggression, and difficulty concentrating, often associated with poor impulse control. Internalizing skills, by contrast, reflect inward-focused behaviors such as anxiety, low self-esteem, and social withdrawal, and are commonly linked to emotional regulation and self-perception. While previous work on rank effects has typically focused on narrow, single-trait outcomes such as self-confidence or academic expectations \parencite{murphy2020, elsner2021}, our use of composite constructs offers a broader and more structured perspective on non-cognitive development.

\vspace{1mm}
We find that a four-position improvement in school-cohort rank — approximately a 10\% increase in relative standing — raises internalizing skills by roughly 4.5\% of a standard deviation, a statistically significant and robust effect across specifications. In contrast, the effect on externalizing skills is smaller, around 3\% of a standard deviation, and less precisely estimated, being statistically significant only at the 10\% level. These average effects conceal notable gender heterogeneity. Boys show gains in both skill dimensions as their rank improves, while girls exhibit a much stronger response in internalizing skills but no meaningful change in externalizing skills. However, the effect on externalizing skills appears to be driven by a small number of extreme cases with severe behavioral issues, as it loses statistical significance once these cases are excluded. In contrast, the internalizing skills effect remains robust even when the most extreme cases are removed, both in the overall sample and when disaggregated by gender. Taken together, these findings suggest that academic rank has a broad but uneven impact on children’s non-cognitive development — influencing self-concept and emotional adjustment more consistently than behavioral regulation. This interpretation is consistent with psychological evidence linking relative academic standing to perceived competence, self-esteem, and long-term expectations \parencite{creemers2013, elsner2021}.

\vspace{1mm}
We find that a four-position improvement in school-cohort rank — around a 10\% increase in relative standing — raises internalizing skills by roughly 4.5\% of a standard deviation, a statistically significant and robust effect across specifications. The effect on externalizing skills is smaller, around 3\% of a standard deviation, and imprecisely estimated (statistically significant at the 10\% level). Importantly, these average effects mask meaningful gender heterogeneity. Boys experience gains in both skill dimensions as their rank improves, while girls show a much stronger response in internalizing skills and no meaningful change in externalizing skills. However, while the effect on externalizing skills is driven by a few extreme cases of poor externalizing behavior, the effect of rank on internalizing skills is robust to the exclusion of poor internalizing students. Even when doing gender heterogeneity, the effect on internalizing skills is robust for both boys and girls, unlike the one on externalizing skills. Taken together, the results suggest that academic rank has a broad but uneven impact on children’s non-cognitive development — shaping self-concept and emotional adjustment more reliably than behavioral regulation. This interpretation aligns with psychological evidence linking relative academic standing to perceived competence, self-esteem, and long-term expectations \parencite{creemers2013, elsner2021}.

\vspace{1mm}
Given the robust impact of rank on non-cognitive development, a natural question is whether these effects arise solely through internal psychological mechanisms or are also shaped by external factors. One potential channel is parental behavior, as parents might respond to their child’s rank by adjusting their level of involvement or support. We find little evidence of systematic responses: there are no consistent effects on parental involvement with homework or on the time children spend studying. The only detectable effect is a modest increase in parental awareness of the teacher’s identity, hinting at some limited effect on their engagement. Overall, the results suggest that rank effects are driven primarily by children's internal adjustments, rather than changes in parental investment.

\vspace{1mm}
To assess the long-term consequences of academic rank, we estimate its effect on adult outcomes using responses from a 2001 follow-up survey, administered nearly 40 years after the original data collection. Despite selective attrition, we show that participation in the follow-up is unrelated to academic rank, allowing for credible identification of long-term effects. Our findings indicate that academic rank in primary school impacts educational attainment. A four-position increase in school-cohort rank raises the probability of attending a grammar school by 4 percentage points — equivalent to a 20\% increase relative to the baseline. The same improvement increases the likelihood of completing O-levels and A-levels (which were standardized qualifications typically taken at ages 16 and 18, respectively, and served as key milestones for educational progression in the UK system) by 5\% and 12\%, respectively. These effects are notably stronger for girls, who appear to benefit more from higher relative rank in terms of access to selective education. Importantly, we do not find an impact of rank on the probability of getting a degree.

\vspace{1mm}
However, these educational gains do not consistently extend to the labor market. We estimate a modest impact on earnings: a four-position rank increase raises the probability of earning more than £25,000 per year by 2.2 percentage points (roughly 6\% of the baseline), and only for boys. There is no effect on higher income thresholds (£35,000 or £45,000), nor on broad measures of socioeconomic status, such as working in high-status occupations. Rank also appears unrelated to family formation: we find no significant effect on the probability of ever marrying or having children. Likewise, there is no detectable impact on subjective well-being, including self-reported happiness and enjoyment of daily activities. On the other hand, higher-ranked children seem to have fonder memories of their time in primary school, as they are more likely to state they were happy at the time.

\vspace{1mm}
Taken together, the results suggest that while early academic rank meaningfully shapes educational trajectories, its influence on economic and personal outcomes is more limited. This pattern is especially striking for girls, whose stronger educational response to rank does not translate into higher earnings or occupational status. These findings likely reflect the historical context of mid-20th-century Scotland, when access to university and professional careers, particularly for women, was far more constrained than today. As a result, rank-driven differences in educational success may have been insufficient to overcome broader structural barriers, limiting the translation of early academic advantage into adult socioeconomic returns.

\vspace{1mm}
We contribute to the growing literature on the effect of academic rank in several ways. We introduce a broader measure of non-cognitive skills, grounded in decades of literature in developmental psychology \parencite{rutter1967, mcgee1985, iloeje1992, klein2009, narusyte2017}. With the exception of \textcite{pagani2021}, previous studies focus on single self-reported traits, such as self-confidence \parencite{murphy2020} or academic expectations \parencite{elsner2017, elsner2021}. Our analysis uses validated composite measures of externalizing and internalizing skills, capturing a richer spectrum of non-cognitive traits (for instance, emotional regulation, impulse control, attention, and social interaction). This approach provides a more comprehensive view of how rank influences children’s behavioral and emotional development, and allows us to test not just whether rank matters for non-cognitive skills, but which dimensions it affects and how robust those effects are. Notably, we show that the rank effect on internalizing skills is both sizable and stable, while the effect on externalizing skills is more fragile and driven by a small subset of students with severe behavioral difficulties.

\vspace{1mm}
We also offer new evidence on the role of parental behavior in mediating the rank effect. The possibility that parents adjust their involvement based on the child’s relative academic standing has been suggested in theory but rarely tested empirically. A recent exception is \textcite{megalokonomou2024}, who show that parents in China respond to their child’s classroom rank by increasing or reducing tutoring investments, interpreting rank as a signal of ability. Our paper complements this work by studying a different type of parental response—day-to-day involvement in the child’s education—using survey data collected from mothers in a randomized subsample. We examine whether rank affects parental knowledge of the child’s teacher, help with homework, and the amount of time children spend studying at home. The results show little systematic evidence of behavioral adjustment beyond a small increase in their awareness of the teacher's name. This suggests that the effects of rank on non-cognitive and academic outcomes may arise more from how children internalize their relative position than from active parental recalibration of effort, especially in lower-stakes educational environments.

\vspace{1mm}
Moreover, we provide rare empirical evidence on the long-term consequences of early academic rank, nearly four decades after the reference point. Few studies beyond \textcite{denning2023}, who link rank to earnings 25 years later, have examined whether early relative standing has persistent effects. We find that academic rank has lasting consequences for educational attainment, particularly for girls, whose probability of attending grammar school and completing formal qualifications responds more strongly to rank than that of boys. These educational gains do not fully translate into higher earnings for women, likely reflecting historical constraints on university access and labor market participation. Overall, by tracing the influence of academic rank across a wide range of adult outcomes — from recalled school experiences to income — we offer a comprehensive account of its long-term relevance.

\vspace{1mm}
Finally, by showing that academic rank influences the development of non-cognitive skills and has long-lasting effects on educational and economic outcomes, our findings contribute to the broader literature on the long-term returns to early skill formation. In particular, they complement evidence that interventions targeting cognitive and non-cognitive development in childhood can yield persistent benefits \parencite{heckman2012, heckman2013, sorrenti2024}, suggesting that relative position in the classroom may be an overlooked yet meaningful mechanism shaping children's life trajectories.

%%%%%%%%
% Data %
%%%%%%%%

\section{Data} \label{sec_2.0}

\vspace{1mm}
This study draws on rich data tracking a cohort of students enrolled in all primary schools in Aberdeen, Scotland, in 1962. Our analysis focuses on children attending grades 3 to 7 at the time, for whom we construct a measure of rank based on performance in a standardized test taken at age 9. We define each student’s rank within their school-cohort group to capture their position in the local academic distribution. We examine the relationship between this early academic rank and a wide set of outcomes measured over the life course. These include academic performance at age 11, non-cognitive skills assessed during primary school, and parental investment for a subset of students whose parents were interviewed. We also study long-term outcomes, including educational attainment, socioeconomic status, fertility, and both physical and mental health, of those who replied to a follow-up survey approximately four decades later, in 2001.

\subsection{The Aberdeen Children of the 1950's Survey} \label{sec_2.1}

\vspace{1mm}
We base our analysis on the Aberdeen Child Development Survey, a comprehensive study conducted between 1962 and 1964.\footnote{Now known as \href{https://web.www.healthdatagateway.org/dataset/4ccb0964-d74a-47b8-96e7-ba5e564c1681}{``The Aberdeen Children of the 1950s''}} Originally designed to investigate the link between anthropometric measures and reading disabilities, the survey was not primarily focused on schooling. Nevertheless, it provides a rich array of information that enables us to track children throughout their primary school years and into adulthood. The dataset includes detailed records of academic progression and achievement, non-cognitive skill measures, physical development indicators, and family background data collected through interviews with a randomly selected subset of parents. In addition, it incorporates follow-up survey data capturing long-term outcomes approximately 40 years later. The data we use are drawn from three distinct sources within the broader study.

\vspace{1mm}
\textbf{The reading survey}: It is the core of the initial data collection effort. Conducted in multiple stages, this survey began in December 1962 with the assistance of teachers, who helped students provide detailed demographic information, including date of birth, father’s occupation, and family size. Schools also contributed attendance records for the preceding two years and available test scores — any missing scores were supplemented in subsequent rounds. In addition, hospital records were used to extract obstetric and social data gathered during the mothers’ pregnancies. By July 1963, further information was collected from schools, including anthropometric measures such as height and weight, taken during routine medical examinations when children entered school (around age 5), and again at ages 9 and 12. The final stage of the Reading Survey took place in March 1964, when teachers completed behavioral assessments. These included the Rutter Questionnaire (scale B) \parencite{rutter1967}, a validated psychological tool for identifying minor behavioral disorders in children.

\vspace{1mm}
\textbf{The family survey}: It targeted a random subsample comprising 25\% of the children included in the Reading Survey. With an overall response rate of 80\%, this follow-up collected information through interviews with the children’s mothers or a substitute caregiver. The survey captured additional insights into the children's health and behavioral conditions, how they allocated their time between leisure and school activities, and provided more detailed demographic information about the parents.

\vspace{1mm}
\textbf{The 2001 follow-up survey}: Participants from the initial cohort were contacted by mail in 2001 and asked to complete a comprehensive questionnaire covering various aspects of their adult lives, including physical and mental health, psychological well-being, and socioeconomic status. Approximately 60\% of those who received the survey returned it with complete responses, allowing us to link early-life information to long-term outcomes.

\subsection{Sample Definition} \label{sec_2.2}

\vspace{1mm}
Our dataset includes 12,151 observations, each corresponding to a child born in Aberdeen between October 1950 and September 1955 and attending grades 3 to 7 at the time. From this initial pool, we apply a series of restrictions to define our analysis sample. First, we exclude students enrolled in private schools (2.88\%) and special schools (9.5\%), as these institutions follow different educational and administrative structures.\footnote{Private schools refused to have their teachers administer the Rutter Questionnaire, which is problematic for our analysis. We do not know enough about how special schools worked to understand if they could be compared to the other public schools in the sample.} We also drop students attending schools that closed before 1964, when the reading survey ended. Finally, we retain only those students for whom we can reliably identify their cohort as of December 1962. After these exclusions, our working sample consists of 9,969 students across 28 schools. For each specific outcome analysis, we further refine the sample to ensure comparability and meaningful inference. When examining the effect of rank on academic performance, we use all students in our working sample. For the analysis of non-cognitive skills, we focus on students still enrolled in primary school as of March 1964, when teachers completed the Rutter Questionnaire. We therefore exclude the cohorts who had already transitioned to secondary or junior secondary school by that time. This restriction leaves us with a sample of 6,779 students.

\subsection{Defining School Cohort in the Scottish Education System} \label{sec_2.3}

\vspace{1mm}
Given that our measure of relative rank is defined within school-cohort groups, it is essential to clarify how we construct cohorts in the context of the institutional setting. In principle, a cohort would correspond to all students enrolling in school during the same academic year, beginning from grade 1 at around age 5. However, the school entry policy in place at the time involved two separate intakes per calendar year: one in January and one in August.\footnote{Based on the statement by the Director of Education to a Town Council of Aberdeen meeting that took place on 3rd October 1960 (\textcite{lawlor2006}) we know that schools had two or more admission dates every academic year. All the students who became five years old before the next admission date had to attend school from the first school day following that admission date.} Children born between October and the end of March were eligible for the January intake, while those born between April and the end of September typically entered school in August. As a result, even within the same grade, students could belong to different intake groups, having started school at different times and progressed separately through the system.

\vspace{1mm}
Because children spent two semesters in each grade before advancing to the next, the two intake streams remained effectively distinct despite students being in the same nominal grade. As a result, children who entered school in different semesters followed parallel but separate academic trajectories. This institutional feature is crucial for our analysis, as it implies that grade alone is not sufficient to identify a student’s cohort. To construct the rank variable accurately, we define each student's cohort based on both the grade they attended in the reference year of the survey (December 1962) and their expected intake group, as inferred from their date of birth. This procedure leaves us with a total of 10 distinct cohorts. We then calculate each student's rank within their school-cohort group, defined by the combination of the school they attended and their inferred cohort.

\subsection{Defining Rank within the School-Cohort Group} \label{sec_2.4}

\vspace{1mm}
We use the standardized outcome of a cognitive skills test taken at age 9 (see \autoref{sec_2.5}) to construct the ranking of students within their school-cohort group. To facilitate comparability across groups of different sizes, we calculate a ``percentilized rank'' following the approach of \textcite{murphy2020}, which normalizes each student’s rank by the total number of students in their group. This approach ensures that the resulting rank measure is bounded between 0 (the lowest-ranked student) and 1 (the highest-ranked student):

\begin{equation} \label{equation_0}
RANK_{isc} = (n_{isc} - 1 )/(N_{sc} - 1)
\end{equation}

\vspace{1mm} \noindent 
Where $n_{isc}$ is the ordinal rank of individual \textit{i} enrolled in school \textit{s}, in cohort \textit{c}. $N\_{sc}$ is the size of the school-cohort group to which the student belongs. This transformation expresses each student’s relative standing as a percentile within their group, ensuring that rank measures are directly comparable across groups of varying sizes. When ties occur, the mean rank is assigned, preserving the average rank of a group and avoiding arbitrary tie-breaking.

\subsection{Standardized Tests} \label{sec_2.5}

\vspace{1mm}
We rely on two standardized tests administered during primary school to measure students' cognitive skills and academic achievement. The first is a low-stakes assessment taken around age 9, which we use to construct relative academic rank and capture baseline cognitive ability. The second, known as the 11-plus test, is a high-stakes examination taken at the end of primary school that played a central role in determining secondary school placement.

\begin{figure} \centering

\caption{Distribution of Cognitive Skills, by Cohort} \label{graph_1}

\includegraphics[scale = 0.80]{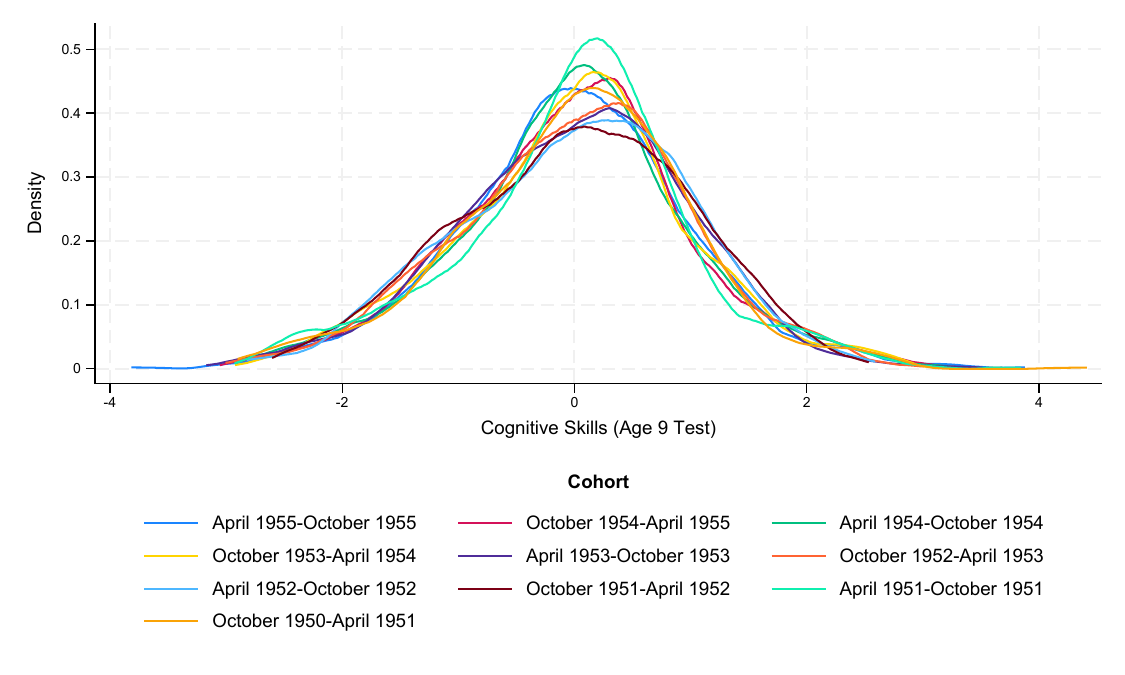}

\begin{minipage}{13 cm} \vspace{1.2ex}
\scriptsize{Notes: We show the distribution of students' cognitive skills, proxied by the outcomes of the Age 9 Test, by cohort. The test scores are standardized at the cohort level. We include only children born between October 1950 and October 1955 and in grades between 3 and 7. The number of observations is 9,368.}
\end{minipage}

\end{figure}

\vspace{1mm}
\textbf{Age 9 Test}: We use performance on the age 9 test as our primary measure of students’ cognitive skills. This test not only serves as the basis for constructing our within-group rank measure, but also offers a reliable baseline assessment of individual academic ability. Its design — focused on verbal and numerical reasoning — was specifically intended to capture core cognitive competencies and to facilitate comparisons of student achievement across different regions. In addition, the test was used as a screening tool for reading difficulties.\footnote{The test was known as the ``Schonell and Adams Essential Intelligence Test (Form B)" \parencite{schonell1940}, and consisted of a battery of 100 questions covering verbal reasoning and arithmetic. It was administered such that children in different cohorts would take it at the same relative age: students who entered school in January were tested in November, while those who started in August were tested in May.} Given the nature and purpose of the test, we refer to the resulting scores as our measure of \textit{cognitive skills} throughout the analysis. Although the test was administered under standardized conditions, the resulting distribution of scores deviates from the standard normal, as shown in \autoref{graph_1}. To assess whether these distributions differ systematically across cohorts, we perform a Kruskal-Wallis test—a non-parametric method that compares the distributions of multiple groups based on the ranks of their values. The results indicate no statistically significant differences in the distributions of cognitive skills across cohorts, suggesting a high degree of comparability in this key measure used to define students’ relative academic rank.

\begin{figure} \centering

\caption{Distribution of the Verbal Reasoning Test, by Cohort} \label{graph_2}

\includegraphics[scale = 0.80]{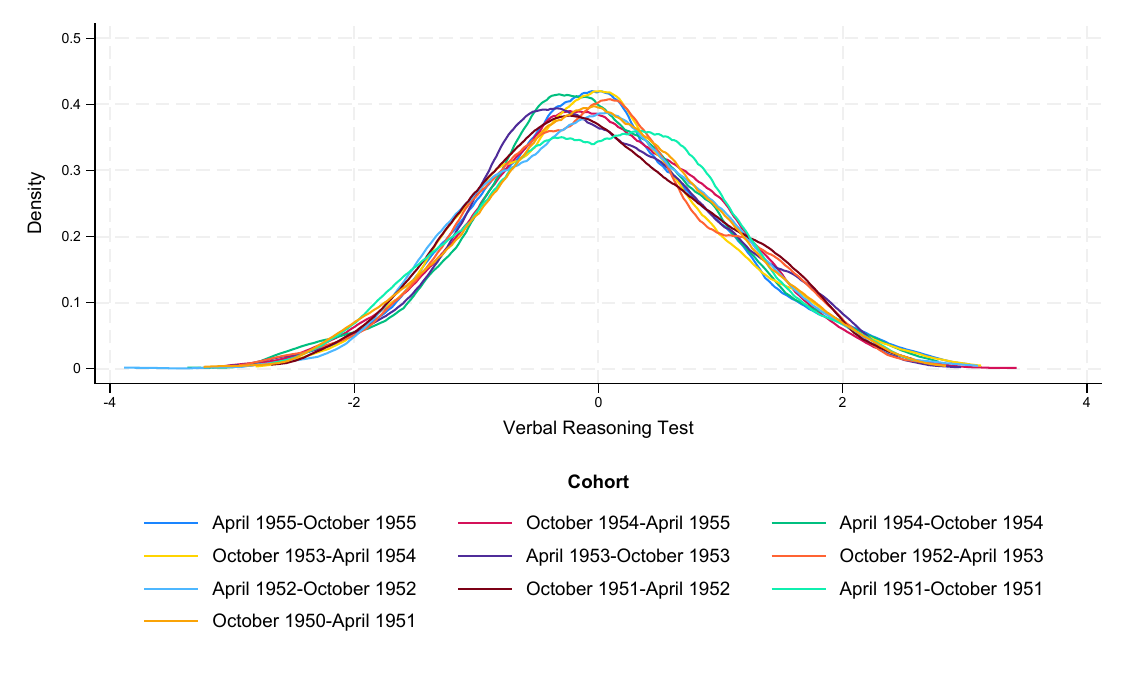}

\begin{minipage}{13 cm} \vspace{1.2ex}
\scriptsize{Notes: We show the distribution of the outcomes of the Verbal Reasoning Test, by cohort. The test scores are standardized at the cohort level. We include only children born between October 1950 and October 1955 and in grades between 3 and 7. The number of observations is 9,698.}
\end{minipage}

\end{figure}

\vspace{1mm}
\textbf{11-plus Test}: Our second measure of academic performance is based on the results of the 11-plus test, a standardized exam administered at the end of primary school to children aged 11 to 12.\footnote{Students sat four separate tests: two verbal reasoning tests, one arithmetic test, and one English test. In addition, teachers provided an estimate of each student’s likely performance, which was scaled to match the mean and standard deviation of the class’s actual results.} This was a high-stakes examination that played a central role in determining admission to grammar schools, with long-term implications for educational attainment and, particularly for girls, labor market outcomes \parencite{clark2016}. Among the components, the two verbal reasoning tests are consistently available for all cohorts, whereas the overall composite score is missing for the two youngest cohorts — those born between October 1954 and October 1955. Consequently, we use verbal reasoning scores as our primary outcome measure, although results using the composite score are also reported where available. Test scores were standardized using both national and Aberdeen-specific norms by year. \autoref{graph_2} illustrates the distribution of standardized verbal reasoning scores across cohorts, which closely resemble a standard normal distribution and show minimal variation. A Kruskal-Wallis test confirms that the score distributions do not differ significantly across cohorts, supporting the comparability of this outcome measure throughout the sample.

%%%%%%%%%%%%%%%%%%%%%%
% Empirical Strategy %
%%%%%%%%%%%%%%%%%%%%%%

\section{Empirical Strategy} \label{sec_3.0}

\vspace{1mm}
We aim to estimate the rank effect on future academic achievements (the 11-plus Test), non-cognitive skills, parental investment, and long-term outcomes. Our empirical strategy is informed by the recently developed literature on the rank effect \parencite{murphy2020, elsner2021, carneiro2023, denning2023}. We start by describing the identification strategy, clarifying our assumptions, and illustrating the identifying variation we exploit.

\subsection{Our Empirical Strategy} \label{sec_3.1}

\vspace{1mm}
We have individual children $i$, enrolled in school $s \in [1, ..., 28]$, in grade $c \in [3, ...,7]$ (as in, all the children born over the 12 month-period which defines grade assignment, hence from October of year $t$ until October of year $t+1$), that enter school with intake $k \in [1,2]$ (where 1 stands for the January intake and 2 stands for the August intake). The group is given by each intake, within a grade, within a school. We refer to the combination of grade and intake as cohorts $c \in [1, ..., 10]$, which we will use as our group unit. The rank of student $i$, enrolled in school $s$, and cohort $c$ within their school-cohort group, $R_{isc}$, is a function of the student's cognitive skills $A_{isc}$ and group characteristics $\overline{W}_{sc}$:

\begin{equation} \label{equation_1} \tag{1}
    R_{isc} = f(A_{isc}, \overline{W}_{sc})
\end{equation}

\vspace{1mm} \noindent
We can estimate the within-cohort rank effect using the following equation:

\begin{equation} \label{equation_2} \tag{2}
    Y_{isc} = \alpha R_{isc} + \beta A_{isc} + \lambda_{s} + \lambda_{c} + \epsilon_{isc}
\end{equation}

\vspace{1mm} \noindent
The variation that allows us to identify the rank effect has two sources:

\begin{enumerate}
    \item [1)] Children with identical cognitive skills, who are in the same school, but belong to different cohorts. 
    \item [2)] Children who are in the same cohort, but are enrolled in different schools.
\end{enumerate}

\vspace{1mm}
For the rank effect to be identified, we need the following conditional independence assumption to hold: $ \mathbbm{E}[\epsilon_{isc}| R_{isc}, A_{isc}, \lambda_{s}, \lambda_{c}] = 0 $. The assumption could fail if the outcomes that we study are correlated with factors that are specific to a group (meaning a school-grade-intake group), or if peer characteristics that affect our outcomes and are different from rank differ across cohorts groups within the same school. We will show that this is not the case in our sample.

\vspace{1mm}
Our final specification adds a few more elements. We want to make sure that the rank effect coefficient is not capturing any higher order peer or ability effect.\footnote{\textcite{denning2023} provide an insightful discussion on this issue. They compare the estimation of the rank effect to an exclusion restriction. Once the common impact of being in a certain group is accounted for, and the groups we are comparing are sufficiently similar, the remaining variation in student outcomes (given their cognitive skills) spurs from peer effects related to rank. While we can corroborate this claim with several robustness checks, it remains an assumption that needs to be discussed on a case-by-case basis.} We add a quadratic polynomial of individual cognitive skills $g(A_{isc})$. Moreover, we include the mean and standard deviation of the group cognitive skills distribution, $\overline{A}_{sc-i}$ and $\sigma_{sc-i}$ (calculated as the mean and standard deviation of the children's test score within their school-cohort group). We also include a vector of individual characteristics $X_{i}$ (which accounts for sex, socioeconomic status, and month of birth). The equation we estimate is the following:

\begin{equation} \label{equation_3} \tag{3}
    Y_{isc} = \alpha R_{isc} + \beta g(A_{isc}) + \gamma_{1} \overline{A}_{sc-i} + \gamma_{2} \sigma_{sc-i} + X_{i}\delta + \lambda_{s} + \lambda_{c} + \epsilon_{isc}    
\end{equation}

\vspace{1mm} \noindent
We include a quadratic polynomials of individual cognitive skills in the specification to allow for a flexible relationship between test scores and later outcomes. This choice is motivated not only by the need to capture higher-order effects, but also to account for potential nonlinearities in how cognitive ability translates into future performance. Omitting this flexibility could lead to misspecification and, consequently, biased estimates of the rank effect. By incorporating polynomials, we reduce the risk that our results are driven by incorrect assumptions about the functional form of this relationship. In addition, we control for the mean and standard deviation of peers' cognitive skills within each school-cohort group, which ensures comparisons are made among students from groups with similar ability profiles — further isolating the contribution of relative rank from variation attributable to group-level cognitive skill differences.

\vspace{1mm}
As we expect the variation in rank to occur at the group level, all our equations are estimated clustering the standard errors at the school-cohort level (or group level).

\subsection{Evidence on the Validity of the Identifying Assumption} \label{sec_3.3}

\vspace{1mm}
Believing that our model can disentangle the rank effect from other types of peer effects is essential, but not sufficient, for our identification strategy to be successful. A key fact we need to establish is that assignment to groups was quasi-random. The month of birth cutoff is insufficient to ensure that is the case. It makes the variation of cognitive skills across cohorts idiosyncratic but does not account for students potentially sorting into schools. The literature separates between active and passive sorting. The former involves children (or, more realistically, parents) selecting their peer group based on their rank preference. The latter implies that children with certain characteristics could be more likely grouped (if, for instance, school choice is non-random, as in our setting). Considering that children are assigned to a certain intake based on their month of birth, we can safely rule out active sorting. On the other hand, passive sorting remains a concern \textcite{denning2023}.

\vspace{1mm}
We use two tests to diagnose whether passive sorting exists in our setting. First, we estimate the relationship between individual characteristics and rank in the school-cohort group. We regress individual characteristics on percentile rank, conditioning on a quadratic polynomial of cognitive skills, $g(A_{isc})$, the mean and standard deviation of the group cognitive skills distribution, $\overline{A}_{sc-i}$ and $\sigma_{sc-i}$, and school and cohort fixed effects. The rank variables and the controls are constructed using our measure of cognitive skills, the outcome of the age 9 test. The dependent variables we use are sex (boy or girl), socioeconomic status (high or low), height and weight at the time of the first medical exam, birth weight, and number of siblings.\footnote{Female is a binary variable equal to one if the child is a girl. Socioeconomic status (SES) is derived from the two-digit occupational code based on ISCO-58; children whose parent holds a position classified between 1 and 20 — corresponding to ``Administrative, Executive, and Managerial Workers'' — are coded as high SES. Height and weight refer to measurements taken during the first school medical exam, typically conducted at school entry. Because children could be examined up to a year apart due to staggered intake and absences, we residualize these measures by age at examination. Specifically, we restrict the sample to children examined within a 12-month age window (dropping 1.5\% of extreme values), regress height and weight separately on a quadratic polynomial in age (in months), and use the residuals as standardized variables. Birth weight is recorded in pounds, as reported in the children's medical records. Number of siblings reflects the total number of siblings living in the household as of December 1962.}

\begin{equation} \label{equation_BC1} \tag{4}
    X_{i} = \alpha R_{isc} + \beta g(A_{isc}) + \gamma_{1} \overline{A}_{sc-i} + \gamma_{2} \sigma_{sc-i} + \lambda_{s} + \lambda_{c} + \epsilon_{isc} 
\end{equation}

\vspace{1mm} \noindent
The top panel of \autoref{table_1} shows the estimated coefficient $\alpha$ for \autoref{equation_BC1}. We do not find any conditional relation between individual characteristics and rank. However, the coefficient for weight is significant at the 10\% level, while the one for height is just barely insignificant. These coefficients are not necessarily problematic, and controlling for the two in our main equation would address the sorting concerns. An additional test can provide more information on whether sorting is happening.

\vspace{1mm}
Because ranking is a function of the distribution of cognitive skills within the group, an association between individual characteristics and features of the distribution of peer cognitive skills $f(\overline{A}_{sc-i})$ can inform us on whether children with certain characteristics are more likely to end up in certain groups. In particular, we compute the mean, standard deviation, and quartiles of the student-specific leave-out peers' outcome in individual cognitive skills. Again, we include a quadratic polynomial of individual cognitive skills, $g(A_{isc})$, and school and cohort fixed effects as controls. 

\begin{equation} \label{equation_BC2} \tag{5}
    X_{i} = \gamma f(\overline{A}_{sc-i}) + \beta g(A_{isc}) + \lambda_{s} + \lambda_{c} + \epsilon_{isc}  
\end{equation}

\vspace{1mm} \noindent
The panels of \autoref{table_1} referring to \autoref{equation_BC2} report estimates of $\gamma$. These results rule out any (conditional) relationship between peer cognitive skills and individual characteristics, suggesting that passive sorting is not an issue in our setting. It corroborates our assumption of conditional quasi-random assignment to school-cohort groups. 

%%%%%%%%%%%%%%%%%%%%%%
% Balancing Exercise %
%%%%%%%%%%%%%%%%%%%%%%

\setlength\extrarowheight{3pt}

\begin{table}[p]
\caption{Balancing Exercise: Individual Characteristics, Rank, and Peer Cognitive Skills} \label{table_1}
\begin{center}
\scalebox{0.7}{

\begin{tabular}{l c c c c c c c} 

    \vspace{-0.35 cm} \\
    \hline
    \vspace{-0.35 cm} \\
    
    Variables & & Woman & High SES & Height & Weight & Birth Weight & Siblings \\

    \vspace{-0.35 cm} \\
    \hline
    \vspace{-0.35 cm} \\

    \multicolumn{8}{c}{\autoref{equation_BC1}: Conditional relation between individual characteristics and rank} \\ 
     
    \vspace{-0.35 cm} \\
    \hline
    \vspace{-0.35 cm} \\
     
     Percentile Rank & & 0.018 & 0.017 & 0.167 & 0.165* & 0.052 & -0.089 \\
     & & (0.045) & (0.027) & (0.105) & (0.096) & (0.097) & (0.063) \\

    \vspace{-0.35 cm} \\
    \hline
    \vspace{-0.35 cm} \\

    \multicolumn{8}{c}{\autoref{equation_BC2}: Conditional relation between individual characteristics and peer quality} \\ 
     
    \vspace{-0.35 cm} \\
    \hline
    \vspace{-0.35 cm} \\
     
     Mean of Peer Cognitive Skills & & -0.0001 & -0.003 & -0.015 & -0.024 & 0.005 & 0.008 \\
     & & (0.007) & (0.004) & (0.023) & (0.018) & (0.017) & (0.011) \\
     
    \vspace{-0.35 cm} \\
    \hline
    \vspace{-0.35 cm} \\
     
     Standard Deviation of Peer Cognitive Skills & & -0.005 & 0.005 & -0.019 & -0.011 & 0.001 & 0.0001 \\
     & & (0.005) & (0.003) & (0.016) & (0.011) & (0.011) & (0.008) \\
     
    \vspace{-0.35 cm} \\
    \hline
    \vspace{-0.35 cm} \\

     25th Percentile of Peer Cognitive Skills & & 0.004 & -0.004 & -0.011 & -0.014 & -0.001 & -0.006 \\
     & & (0.007) & (0.004) & (0.022) & (0.016) & (0.015) & (0.010) \\
     
    \vspace{-0.35 cm} \\
    \hline
    \vspace{-0.35 cm} \\

     50th Percentile of Peer Cognitive Skills & & -0.004 & -0.004 & 0.001 & -0.001 & 0.020 & 0.010 \\
     & & (0.006) & (0.003) & (0.020) & (0.015) & (0.015) & (0.010) \\
     
    \vspace{-0.35 cm} \\
    \hline
    \vspace{-0.35 cm} \\

     75th Percentile of Peer Cognitive Skills & & 0.001 & -0.001 & -0.010 & -0.017 & 0.006 & 0.010 \\
     & & (0.006) & (0.003) & (0.016) & (0.013) & (0.015) & (0.010) \\
     
    \vspace{-0.35 cm} \\
    \hline
    \vspace{-0.35 cm} \\

    Observations & & 9,698 & 9,698 & 9,465 & 9,458 & 9,698 & 9,698 \\

    \vspace{-0.35 cm} \\
    \hline
    \vspace{-0.35 cm} \\
    
    \end{tabular}
    }

\begin{minipage}{13 cm} \vspace{1.2ex}
\scriptsize{Notes: We estimate the relationship between ranking (\autoref{equation_BC1})/peer cognitive skills (\autoref{equation_BC2}) at the school-cohort group level on different characteristics of the students. These characteristics are: the student probability of being a girl, the student probability of coming from an advantaged socioeconomic background (based on the father's occupation), the student height and weight at the time of their first medical exam, the student birth weight (lbs), and the student number of siblings. We include all 10 cohorts of children who attended primary school in Aberdeen in December 1962, who were born between October 1950 and October 1955. Standard errors are clustered at the school-cohort-group level. *** p $<$ 0.01, ** p $<$ 0.05, * p $<$ 0.1.}
\end{minipage}

\end{center}
\end{table}

%%%%%%%%%%%%%%%%%%%%%%%%%%%%%
% Non-cognitive Development %
%%%%%%%%%%%%%%%%%%%%%%%%%%%%%

\section{Non-Cognitive Development: Externalizing and Internalizing Skills} \label{sec_4.0}

\vspace{1mm}
We measure non-cognitive skills using the Rutter Children’s Behaviour Questionnaire \parencite{rutter1967}, which was completed by teachers in March 1964 for each child in their classroom. This well-established behavioral assessment tool consists of 26 items designed to detect minor behavioral issues in children and has been widely adopted in educational and clinical research settings \parencite{behar1974, boyle1985, mcgee1985, iloeje1992, klein2009, narusyte2017}. Teachers were asked to rate each behavior as “Does not apply,” “Somewhat applies,” or “Definitely applies” for each child, providing a structured and consistent measure of classroom behavior.\footnote{The 26 items in the questionnaire include: ``Restless'', ``Truant'', ``Fidgety'', ``Destroys Belongings'', ``Fights other Children'', ``Disliked'', ``Anxious'', ``Solitary'', ``Irritable'', ``Often Unhappy'', ``Tics'', ``Sucks Fingers'', ``Nail Biting'', ``Trivial Absences'', ``Disobedient'', ``Short Attention Span'', ``Fearful'', ``Fussy'', ``Often Lies'', ``Stealing'', ``Wet/Soiled Themselves'', ``Often Aching/in Pain'', ``Tearful'', ``Stutters'', ``Other Speech Difficulty'', ``Bullies other Children''.}

\vspace{1mm}
The 26 behaviors captured in the Rutter scale refer to two broad domains: externalizing and internalizing behaviors, following a well-established classification in child psychiatry \parencite{achenbach1978, eisenberg2001}. Externalizing behaviors are outward-directed and often reflect poor self-regulation—such as inattention, impulsivity, or aggression. Internalizing behaviors, by contrast, are inward-directed and typically relate to emotional states such as anxiety, sadness, or social withdrawal. These two dimensions provide distinct but complementary perspectives on a child's non-cognitive functioning and social-emotional development.

\vspace{1mm}
Similarly to \textcite{attanasio2020}, we interpret these behavioral traits in terms of skills rather than symptoms. Externalizing skills reflect a child’s ability to regulate impulses, sustain attention, and interact appropriately with others — skills crucial for classroom engagement and cooperation. Internalizing skills, on the other hand, relate to the capacity to channel focus and emotional awareness toward task performance, including self-discipline and self-awareness. Each item on the Rutter scale is scored such that higher values indicate more positive behavioral traits: a score of 0 corresponds to “Definitely Applies,” 1 to “Somewhat Applies,” and 2 to “Does Not Apply.” This scoring convention allows us to interpret higher scores — and positive coefficients in our analysis — as indicating stronger non-cognitive skills.

\subsection{Extracting Measures of Non-Cognitive Skills: Factor Analysis} \label{sec_4.1}

\vspace{1mm}
We rely on principal (common) factor analysis to isolate two latent constructs underlying our behavioral data: externalizing and internalizing skills. This statistical technique captures the shared variation across multiple observed variables — here, questionnaire items — and estimates scores for unobserved (latent) traits that best explain that variation. Given the ordinal nature of the items in the Rutter scale, we use a polychoric correlation matrix to obtain consistent and unbiased estimates of the relationships between items \parencite{olsson1979}. To determine the appropriate number of factors, we combine theoretical guidance from the psychology literature \parencite{behar1974, boyle1985, mcgee1985, iloeje1992, klein2009, narusyte2017} with empirical criteria.

\begin{figure} \centering

\caption{Distribution of Externalizing and Internalizing Skills, by Cohort} \label{graph_3A}

\includegraphics[scale = 0.80]{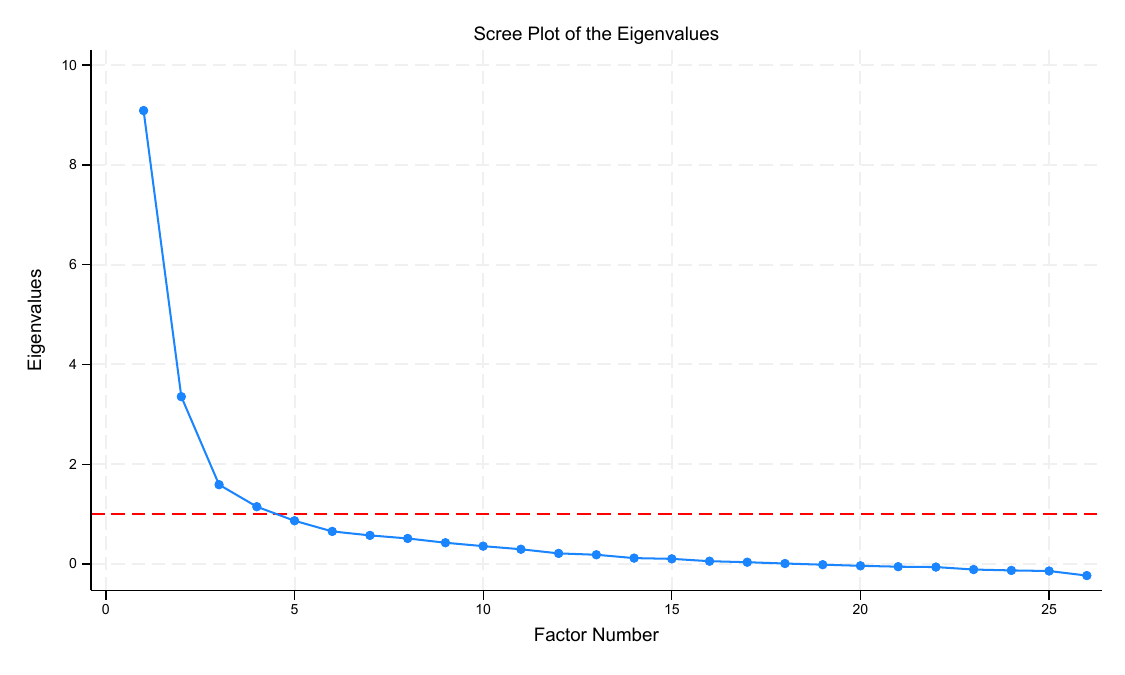}

\begin{minipage}{13 cm} \vspace{1.2ex}
\scriptsize{Notes: The graph plots the eigenvalue of each factor estimated through the first iteration of factor analysis. The number of observations is 6,779.}
\end{minipage}

\end{figure}

\vspace{1mm}
The scree plot of eigenvalues from the initial factor extraction helps identify the number of relevant factors. Following the rule proposed by \textcite{kaiser1960}, we retain factors with eigenvalues greater than one, indicating they explain more variance than any individual item. As shown in \autoref{graph_3A}, four factors exceed this threshold, but two clearly dominate in terms of explained variance — consistent with our theoretical expectation of a two-factor structure.

\vspace{1mm}
We then refine the item set used to define each factor. This step is based on rotated factor loadings and item-specific uniqueness values, obtained through oblique (quartimin) rotation, which allows for correlation between factors and improves interpretability. We exclude items that exhibit weak association with any factor (factor loading below 0.4) or high idiosyncratic variance (uniqueness above 0.8). \autoref{table_2} presents the results from both the initial and refined iterations. After dropping non-informative items, 19 remain and load clearly onto one of the two factors, corresponding to externalizing or internalizing skills. A second round of factor analysis confirms the stability of this structure. Together, the two retained factors explain approximately 80\% of the total variance in the selected items.

\setlength\extrarowheight{3pt}

\begin{table}[p]
\caption{Rotated Factor Loadings from the Exploratory Factor Analysis based on the 26 items of the Rutter Questionnaire for Teachers} \label{table_2}
\begin{center}
\scalebox{0.7}{
\begin{tabular}[t]{l c c | c c}

    \vspace{-0.35 cm} \\
    \hline
    \vspace{-0.35 cm} \\
    
     & \multicolumn{2}{c}{Iteration 1} & \multicolumn{2}{c}{Iteration 2} \\
    Item & Externalizing & Internalizing & Externalizing & Internalizing \\ 

    \vspace{-0.35 cm} \\
    \hline
    \vspace{-0.35 cm} \\
    
    Restless & \textbf{0.78} & 0.00 & \textbf{0.79} & -0.02 \\ 
    Truant & \textbf{0.68} & 0.18 & \textbf{0.67} & 0.16 \\ 
    Fidgety & \textbf{0.77} & 0.01 & \textbf{0.76} & -0.01 \\
    Destroys Belongings & \textbf{0.89} & -0.07 & \textbf{0.89} & -0.07 \\
    Fights Others & \textbf{0.87} & -0.04 & \textbf{0.88} & -0.004 \\ 
    Disliked & \textbf{0.67} & 0.33 & \textbf{0.68} & 0.33  \\ 
    Anxious & -0.16 & \textbf{0.85} & -0.15 & \textbf{0.86} \\
    Solitary & 0.11 & \textbf{0.62} & 0.12 & \textbf{0.60}  \\ 
    Irritable & \textbf{0.75} & 0.04 & \textbf{0.76} & 0.06 \\ 
    Often Unhappy and Miserable & 0.22 & \textbf{0.75} & 0.24 & \textbf{0.76} \\ 
    Tics & 0.39 & 0.32 & - & - \\ 
    Sucks Finger & 0.26 & 0.25 & - & - \\ 
    Nail Biting & 0.24 & 0.13 & - & - \\ 
    Trivial Absences & 0.38 & 0.34 & - & - \\ 
    Disobedient & \textbf{0.87} & -0.12 & \textbf{0.87} & -0.11 \\
    Poor Concentration & \textbf{0.57} & 0.24 & \textbf{0.56} & 0.20 \\
    Afraid & -0.14 & \textbf{0.85} & -0.12 & \textbf{0.84} \\
    Fussy over particular child & -0.18 & \textbf{0.55} & -0.16 & \textbf{0.58} \\
    Often Lies & \textbf{0.86} & 0.004 & \textbf{0.86} & 0.01 \\
    Stealing & \textbf{0.71} & -0.02 & \textbf{0.70} & 0.003 \\
    Wet/Soiled Themselves & 0.26 & 0.29 & - & - \\
    Often Aching & 0.16 & \textbf{0.53} & 0.17 & \textbf{0.49} \\
    Tearful & 0.20 & \textbf{0.63} & 0.21 &\textbf{0.65} \\
    Stutters & 0.20 & 0.29 & - & - \\
    Speech Difficulties & 0.19 & 0.21 & - & - \\
    Bullies Others & \textbf{0.85} & -0.11 & \textbf{0.85} & -0.09  \\
    
    \vspace{-0.35 cm} \\
    \hline
    \vspace{-0.35 cm} \\
    
    \end{tabular}
    }
\begin{minipage}{13 cm} \vspace{1.2ex}
\scriptsize{Notes: We iterate exploratory factor analysis to decide which items to retain out of the 26 in the Rutter Questionnaire for Teachers. We report the factor loadings and the communities for the oblique rotated total variance matrix. We restrict our sample to children born between April 1952 and October 1955, since we want to include only children who were in primary school when the Rutter Questionnaire was completed (March 1964). In total, we have 6,779 children.}
\end{minipage}
\end{center}
\end{table}

\vspace{1mm}
To compute individual-level scores for each skill dimension, we apply the Bartlett method \parencite{hershberger2005}, which uses maximum likelihood estimation to generate unbiased factor scores. Each child receives a score for externalizing and internalizing skills, which we standardize by cohort to ensure comparability across groups. The reliability of the two indices, as measured by Cronbach’s alpha — a statistic that captures internal consistency among items in a scale — is 0.84 for externalizing skills and 0.66 for internalizing skills. These values suggest high reliability for the former and moderate reliability for the latter.

\begin{figure} \centering

\caption{Distribution of Externalizing and Internalizing Skills, by Cohort} \label{graph_3B}

\includegraphics[scale = 0.54]{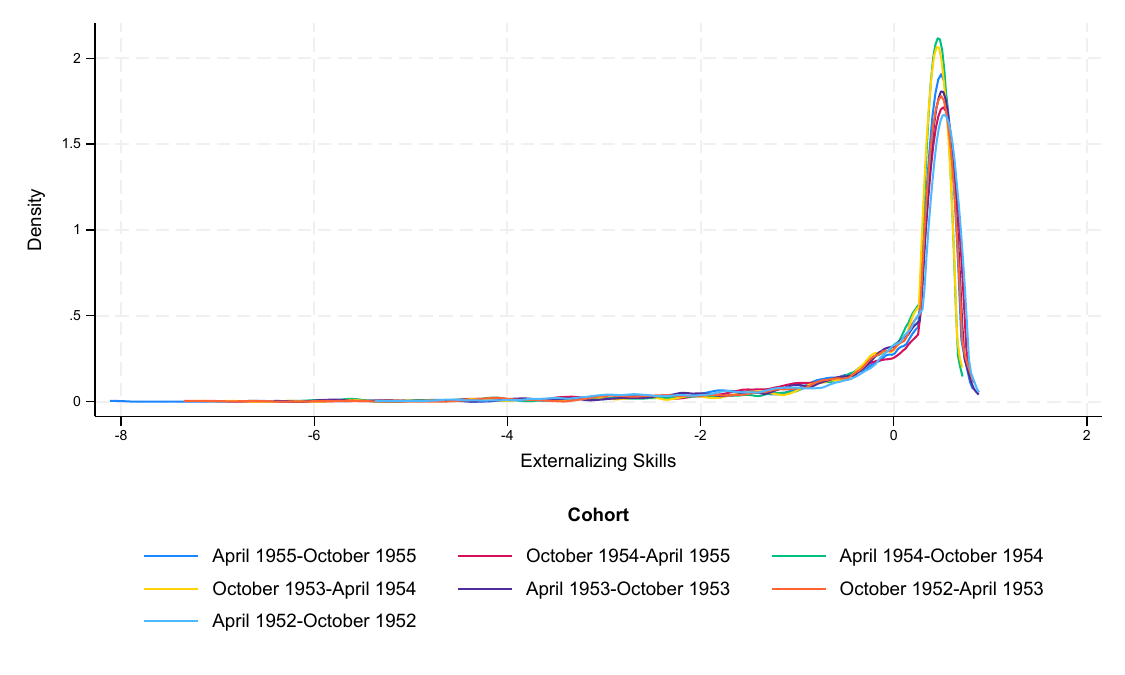}
\includegraphics[scale = 0.54]{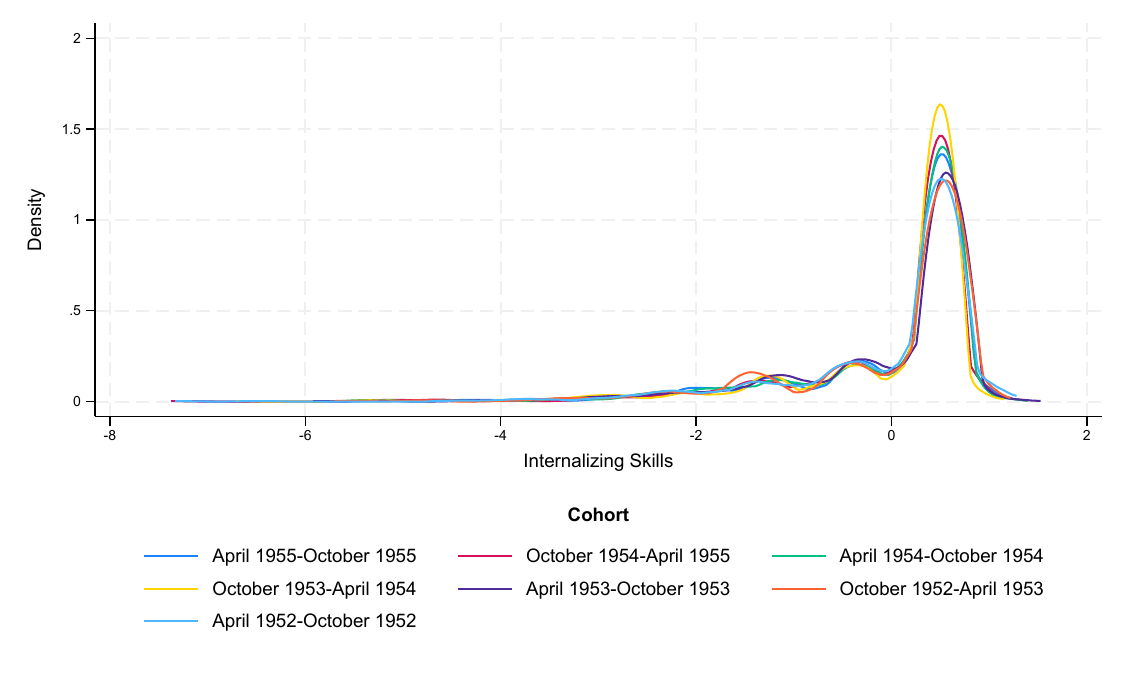}

\begin{minipage}{13 cm} \vspace{1.2ex}
\scriptsize{Notes: The graphs show the distribution of externalizing skills in the top panel and internalizing skills in the bottom panel, by cohort. The variables are standardized at the cohort level. We include only children born between October 1951 and October 1955 and in grades between 3 and 6. The number of observations is 6,779.}
\end{minipage}

\end{figure}

\vspace{1mm}
\autoref{graph_3B} presents the standardized distributions of the two skill measures by cohort. Externalizing skills are shown on the left-hand side, while internalizing skills are displayed on the right-hand side. As expected for an index originally developed to identify mild behavioral difficulties, both distributions are fairly concentrated, with most children clustered in a narrow range. Each distribution displays a long left tail, representing a smaller share of children with more pronounced behavioral challenges. A Kruskal-Wallis test confirms that the distributions of both skill measures differ significantly across cohorts. This likely reflects differences in age at the time of assessment, as children in different cohorts were tested simultaneously in March 1964, but varied in age depending on their intake group. These differences underscore the importance of standardizing scores by cohort in all subsequent analyses.

\subsection{Non-Cognitive Skills within the Rank Effect Literature} \label{sec_4.2}

\vspace{1mm}
A growing body of research has explored how a student's relative position within their peer group - usually school-cohort or classroom - affects non-cognitive development. Existing studies often focus on specific traits such as conscientiousness, self-esteem, self-confidence, or expectations, typically using narrow constructs to test theoretical mechanisms through which rank may operate. While this work provides important insights, it often considers individual traits in isolation, potentially missing the broader impact rank may have across a range of non-cognitive dimensions.

\vspace{1mm}
Our approach broadens this perspective by using two well-established constructs from developmental psychology: externalizing and internalizing skills. These composite measures capture clusters of behaviors that reflect children’s ability to regulate themselves (externalizing skills) and to form a stable self-concept and emotional understanding (internalizing skills). This structure allows us to consider how rank affects a more complete set of behaviors relevant to learning, social interaction, and self-perception, without committing to a single, tightly defined mechanism.

\vspace{1mm}
Several strands of the literature suggest that both types of skills may respond to changes in classroom rank. For instance, improvements in externalizing skills may reflect better impulse control or attentional focus among students who feel pressure to maintain a high position in the classroom hierarchy — a pattern consistent with findings on conscientiousness \parencite{pagani2021}. Conversely, internalizing skills may respond to rank through shifts in self-confidence or perceived academic ability, consistent with work showing that students' expectations, self-beliefs, and aspirations are shaped by their standing relative to peers \parencite{murphy2020, elsner2017, elsner2021}. Moreover, psychological evidence links lower internalizing skill development to damaged self-esteem \parencite{creemers2013}, suggesting a plausible pathway through which rank may influence emotional well-being.

\vspace{1mm}
By studying the effect of academic rank on these two broader skill dimensions, we provide a more holistic view of how the learning environments shape non-cognitive development. While narrow constructs such as self-confidence or conscientiousness offer clarity in testing specific mechanisms, they may overlook the fact that relative academic standing likely influences multiple, interrelated aspects of a child's psychological development. Our contribution is to move beyond these single-trait approaches, using externalizing and internalizing skills to trace the wider non-cognitive implications of being ranked among peers.

\subsection{Adapted Specification for the Estimation of the Rank Effect on Non-Cognitive Skills} \label{sec_4.3}

\vspace{1mm}
A key challenge in estimating the effect of academic rank on non-cognitive skills is the lack of a baseline measure of non-cognitive traits prior to the formation of rank. While the quasi-random assignment of students to school-cohort groups mitigates concerns about selection into rank, controlling only for cognitive skills may not be sufficient. If non-cognitive traits at school entry are correlated with, or causally related to, early academic performance, omitting them could bias the estimated rank effect.

\vspace{1mm}
To address this, we turn to insights from developmental psychology. The literature identifies three interrelated dimensions of child development: cognitive, non-cognitive, and physical \parencite{santrock2020, berk2023}. While difficult to fully disentangle, these domains are strongly correlated. We leverage this correlation by using available proxies from the cognitive and physical domains as baseline controls for the unobserved non-cognitive traits at school entry. In particular, \textcite{duckworth2019} show that physical development is often more closely related to non-cognitive skills than cognitive ability is, further motivating our use of anthropometric indicators.

\vspace{1mm}
Guided by this evidence, we augment our baseline specification (\autoref{equation_3}) with two additional controls capturing early physical development. The first, $j(H_{isc})$ is a quadratic polynomial in residualized height measured at the first school medical exam (as explained in \autoref{sec_3.3}). The second, $k(B_{isc})$, is a quadratic polynomial in birth weight, measured in pounds, provided by the children's medical records. The resulting specification is:

\begin{equation} \label{equation_F2} \tag{6}
    Y_{isc} = \alpha R_{isc} + \beta g(A_{isc}) + \gamma_{1} \overline{A}_{sc-i} + \gamma_{2} \sigma_{sc-i} + \theta_{1} j(H_{isc}) + \theta_{2} k(B_{isc}) + X_{i}\delta + \lambda_{s} + \lambda_{c} + \epsilon_{isc}    
\end{equation}

\vspace{1mm}
Here, $Y_{isc}$ is the standardized score for either externalizing or internalizing skills, $R_{isc}$ denotes a student’s relative rank, and $g(A_{isc})$ is a quadratic polynomial in individual cognitive ability. $\overline{A}_{sc-i}$ and $\sigma_{sc-i}$ are, respectively, the mean and standard deviation of cognitive skills within a school-cohort group, excluding the individual. $j(H_{isc})$ and $k(B_{isc})$ capture physical development. $X_{i}$ includes additional student-level controls, while $\lambda_{s}$ and $\lambda_{c}$ are school and cohort fixed effects, respectively.

%%%%%%%%%%%
% Results %
%%%%%%%%%%%

\section{Results} \label{sec_5.0}

\vspace{1mm}
We now turn to the empirical analysis of the effects of academic rank. Using our measure of relative standing within school-cohort groups—constructed from students’ performance on the age 9 test — we estimate the impact of rank on a wide range of outcomes. These include short-run academic achievement, non-cognitive skill development, parental investment, and long-term educational, economic, and health-related outcomes. By relying on variation in rank that is orthogonal to cognitive skills, our analysis isolates the effect of a student's position in the local academic distribution on later-life trajectories.

\subsection{The Rank Effect on Academic Performance} \label{sec_5.1}

\vspace{1mm}
We estimate the effect of academic rank on students' academic performance, proxied by results from the high-stakes 11-plus examination. As discussed, this test played a central role in determining secondary school placement and had lasting consequences for educational and labor market outcomes \parencite{clark2016}. In \autoref{table_3}, we present the results. Columns (1) and (2) report estimates using verbal reasoning test (VRT) scores, which are available for all the cohorts in our sample. Columns (3) and (4) show estimates using the overall 11-plus composite score, which is unavailable for the two youngest cohorts.

\vspace{1mm}
Across all specifications, we find a positive and statistically significant effect of rank on academic performance. The magnitude of the estimated coefficient is remarkably similar whether we use the VRT or the overall 11-plus score, with only a minimal decline observed when moving to the composite measure. Moreover, controlling for individual characteristics—including gender, socioeconomic status, height, weight, birth weight, and number of siblings—does not substantially alter the estimates, as seen from the comparison between columns (1) and (2) and between columns (3) and (4).

\vspace{1mm}
The estimated coefficient of approximately 0.55 implies that, holding cognitive skills and other characteristics constant, a student ranked at the top of their school-cohort group would perform about 60\% of a standard deviation better on the 11-plus test than a student ranked at the bottom. A more granular interpretation emerges if we consider the effect of a 10\% increase in relative rank. Given that the average school-cohort group size is 37 students (with a median of 29), moving up roughly 4 positions within the group is associated with an increase of 0.055 standard deviations in test scores, equivalent to about 6\% of a standard deviation.

\vspace{1mm}
Our findings are consistent with previous estimates in the literature. \textcite{murphy2020} document similar magnitudes when examining the impact of relative rank on academic achievement in English primary schools, while \textcite{elsner2021} find comparable effects using data from German secondary schools. These results reinforce the conclusion that a student's position in the local academic distribution has meaningful consequences for measured academic success, even after controlling for absolute ability.

\begin{figure} \centering

\caption{Rank effect on the (standardized) outcome of the Verbal Reasoning and 11-plus tests, by rank decile} \label{graph_4}

\includegraphics[scale = 0.54]{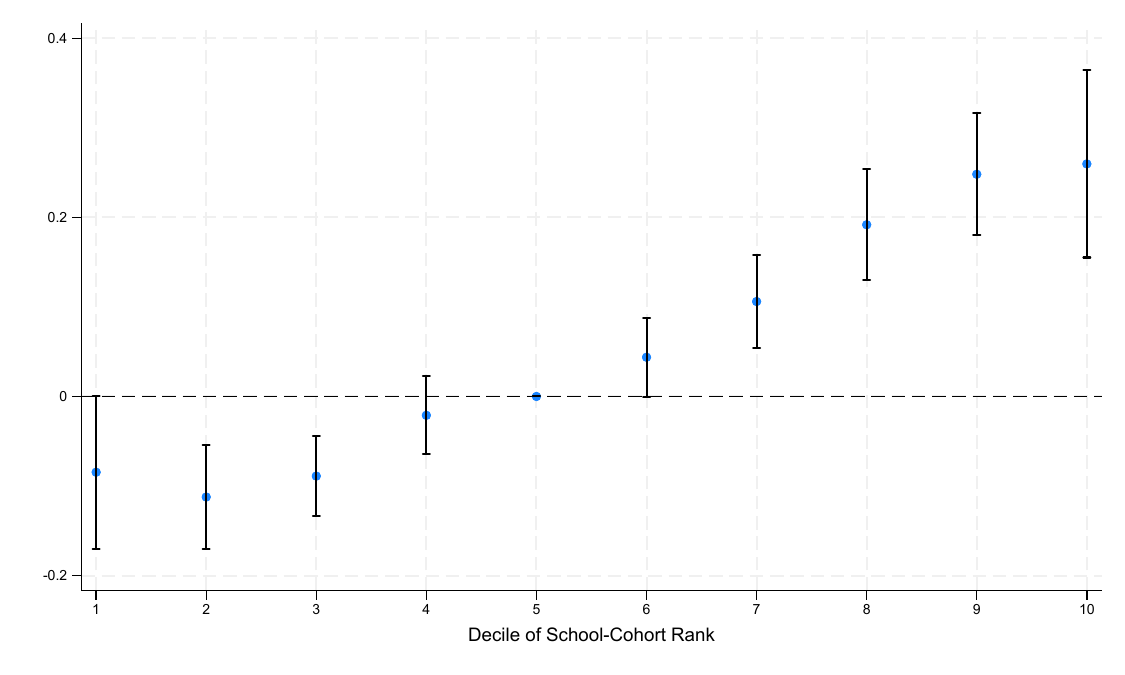}
\includegraphics[scale = 0.54]{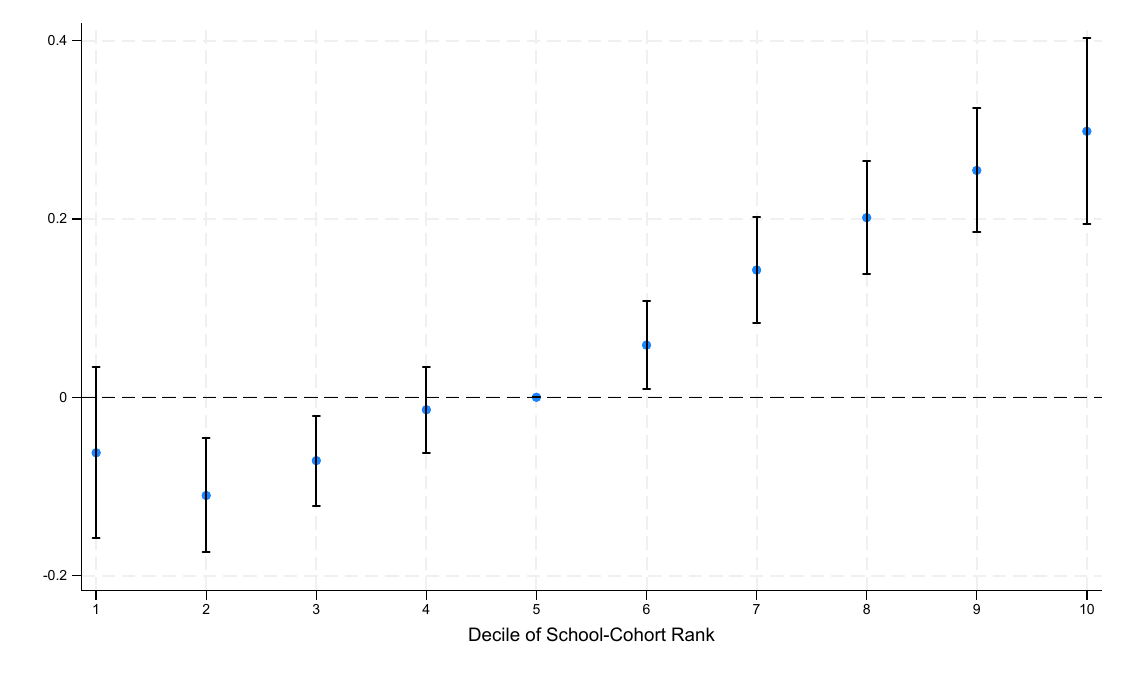}

\begin{minipage}{13 cm} \vspace{1.2ex}
\scriptsize{Notes: The graph shows the rank effect on the Verbal Reasoning and the 11-plus test scores by rank decile.}
\end{minipage}

\end{figure}

\vspace{1mm}
We assume that the effect of academic rank on outcomes is linear and control for potential non-linearities in cognitive skills by including a quadratic polynomial. To assess the plausibility of this assumption, we break down the rank distribution into deciles and plot the estimated effects in \autoref{graph_4}. The relationship appears broadly linear across the distribution, for the effects on both the VRT and the 11-plus test, supporting our baseline specification. These patterns also support the self-reinforcing nature of the rank effect: early differences in rank lead to performance gains that influence future rank, making the timing of cognitive skill measurement less critical.

%%%%%%%%%%%%%%%%%%%%%%%%%%%%%%%%%
% Results: Academic Performance %
%%%%%%%%%%%%%%%%%%%%%%%%%%%%%%%%%

\setlength\extrarowheight{3pt}

\begin{table}[p]
\caption{Rank Effect on the 11-plus Test} \label{table_3}
\begin{center}
\scalebox{0.7}{
\begin{tabular}{l c c c c}
    
    \vspace{-0.35 cm} \\
    \hline
    \vspace{-0.35 cm} \\

                      & (1) & (2) & (3) & (4) \\
    Outcome Variables & \multicolumn{2}{c}{Verbal Reasoning Test} & \multicolumn{2}{c}{11-plus Test} \\
        
    \vspace{-0.35 cm} \\
    \hline
    \vspace{-0.35 cm} \\
        
    Percentile Rank & 0.575*** & 0.579*** & 0.540*** & 0.544*** \\
                    & (0.068)  & (0.068)  & (0.074)  & (0.074)  \\
     
    \\

    Mean of the Outcome & 0     & 0     & 0     & 0     \\
    SD of the Outcome   & 1     & 1     & 1     & 1     \\
    Observations        & 9,441 & 9,441 & 7,575 & 7,575 \\
        
    \vspace{-0.35 cm} \\
    \hline
    \vspace{-0.35 cm} \\

    School Fixed Effects & X & X & X & X \\
    Cohort Fixed Effects & X & X & X & X \\
    
    \\
    
    Cognitive Skills         & X & X & X & X \\
    Cognitive Skills Squared & X & X & X & X \\

    \\

    Mean of Peer Cognitive Skills & X & X & X & X \\
    SD of Peer Cognitive Skills   & X & X & X & X \\
    
    \\
    
    Sex                  & - & X & - & X \\
    Socioeconomic Status & - & X & - & X \\
    Number of Siblings   & - & X & - & X \\
    Month of Birth       & - & X & - & X \\
    
    \vspace{-0.35 cm} \\
    \hline
    \vspace{-0.35 cm} \\

\end{tabular}
}
\begin{minipage}{13 cm} \vspace{1.2ex}
\scriptsize{Notes: We estimate the relationship percentile rank within the school-cohort group and the standardized outcome of the Verbal Reasoning Test, as well as the standardized outcome of the 11-plus Test. Percentile rank is established using our baseline measure of cognitive skills, the outcome of the Age 9 Test. The 11-plus Test consists of 4 components: two Verbal Reasoning Tests, one Algebra Test, and one English Test. We consider the outcome of the Verbal Reasoning Tests first, as it is available for all 10 cohorts of children in our survey; the overall outcome of the 11-plus test is not available for the two youngest cohorts of children in our sample. Our sample consists of 10 cohorts of children who attended primary school in Aberdeen in December 1962, who were born between October 1950 and October 1955. We control for: a categorical variable taking a different value for each school in the sample; a categorical variable taking a different value for each cohort in the sample; a quadratic polynomial of child cognitive skills (based on the outcome of the Age 9 Test); the mean and standard deviation of the cognitive skills of the peers of the students (based on the outcome of the Age 9 Test); a categorical variable taking value 1 if the child is a girl, and 0 if he is a boy; a categorical variable taking value 1 if the child belongs to a family of high socioeconomic status (defined based on the father's occupation), and 0 otherwise; a categorical variable capturing the specific month of birth of the child; and the number of siblings of the child. Standard errors are clustered at the school-cohort level.}
\end{minipage}
\end{center}
\end{table} 

\subsection{The Rank Effect on Non-Cognitive Skills} \label{sec_5.2}

\vspace{1mm}
We next estimate the impact of academic rank on non-cognitive skills, focusing on externalizing and internalizing skills as derived from the Rutter Questionnaire. Rather than capturing a single narrow trait, these measures bundle together a set of behaviors relevant to self-regulation, social interaction, emotional stability, and self-perception, offering a broad perspective on children's non-cognitive development.

\vspace{1mm}
The estimates are reported in \autoref{table_4A}. In columns (1) and (4), we regress externalizing and internalizing skills, respectively, on school-cohort rank, controlling for school and cohort fixed effects, a quadratic polynomial of individual cognitive skills, and the mean and standard deviation of peers' cognitive skills within the school-cohort group. In columns (2) and (5), we further add controls for individual characteristics — sex, socioeconomic status, number of siblings, and month of birth — to account for potential differences in family background and demographic factors. Finally, in columns (3) and (6), we also control for early physical development through height, weight, and birth weight, as these may correlate with non-cognitive traits at school entry and thus strengthen the robustness of our estimates. As stated in \autoref{sec_2.2}, we include only children who were still in primary school as of March 1964,  when the teachers completed the Rutter Test's questionnaire (therefore, those in the 7 youngest cohorts, born between April 1952 and October 1955).

\vspace{1mm}
We find a positive and relatively stable effect of academic rank on externalizing skills across specifications, although the estimates are imprecisely measured and statistically significant only at the 10\% level. A four-position increase within the school-cohort group (approximately 10\% of the average number of students in the group, which is 37) is associated with an improvement in externalizing skills of around 3\% of a standard deviation. In contrast, the effect on internalizing skills is larger and estimated with greater precision: the same 10\% jump in relative rank corresponds to an increase of approximately 4.5\% of a standard deviation in internalizing skills.

\vspace{1mm}
As shown in \autoref{graph_3B}, the distributions of externalizing and internalizing skills exhibit long left tails, reflecting a small number of children with severe behavioral difficulties. To ensure that our estimated rank effects are not disproportionately driven by these extreme cases, we re-estimate our main specifications after excluding the bottom 3\% and 5\% of each distribution. The results, reported in \autoref{table_4B}, reveal a clear contrast between the two skill dimensions. The effect of rank on externalizing skills is highly sensitive to these exclusions. Removing just the bottom 3\% of children leads to a sharp drop in the estimated coefficient and a complete loss of statistical significance. By contrast, the effect on internalizing skills remains robust, with only modest reductions in magnitude. A four-position improvement in rank continues to correspond to an increase of roughly 3.7\% of a standard deviation, even when the most severe 5\% of cases are excluded, compared to about 4.9\% in the full sample. This pattern holds even when disaggregating by gender (see \autoref{sec_5.5}), with the effect for boys — who primarily drove the overall estimate — also becoming insignificant. That does not happen to internalizing skills, which maintain their significant effect for both boys and girls. This indicates that the rank effect on internalizing skills is more robust and generalizable across the broader population of students.\footnote{All our main results, including those on academic performance, long-term educational attainment, and adult outcomes, remain statistically significant and qualitatively similar when excluding the most extreme 3\% to 5\% of cases in the non-cognitive skills distribution, indicating that our findings are not driven by a small subset of extremely problematic students.}

\vspace{1mm}
Our findings fit naturally within the emerging literature documenting the influence of relative academic standing on non-cognitive development. Prior studies have tended to focus on single traits such as self-esteem, conscientiousness, or expectations \parencite[e.g.,][]{murphy2020, pagani2021, elsner2021}. By employing composite measures of externalizing and internalizing skills, we provide a more comprehensive picture of how rank shapes multiple dimensions of children's psychological and behavioral development. While our results do not point to a single definitive mechanism, the stronger and more consistent effect on internalizing skills suggests that rank may primarily operate by influencing children's self-concept and emotional self-perception, rather than solely affecting behaviors related to attention or impulse control. This interpretation is consistent with existing evidence that academic rank affects self-confidence, self-esteem, and educational expectations \parencite{murphy2020, elsner2017, elsner2021}. In this sense, the rank effect appears to shape not only how students behave, but also how they view themselves within the educational environment.

\begin{figure} \centering

\caption{Rank effect on (standardized) externalizing and internalizing skills, by rank decile} \label{graph_5}

\includegraphics[scale = 0.54]{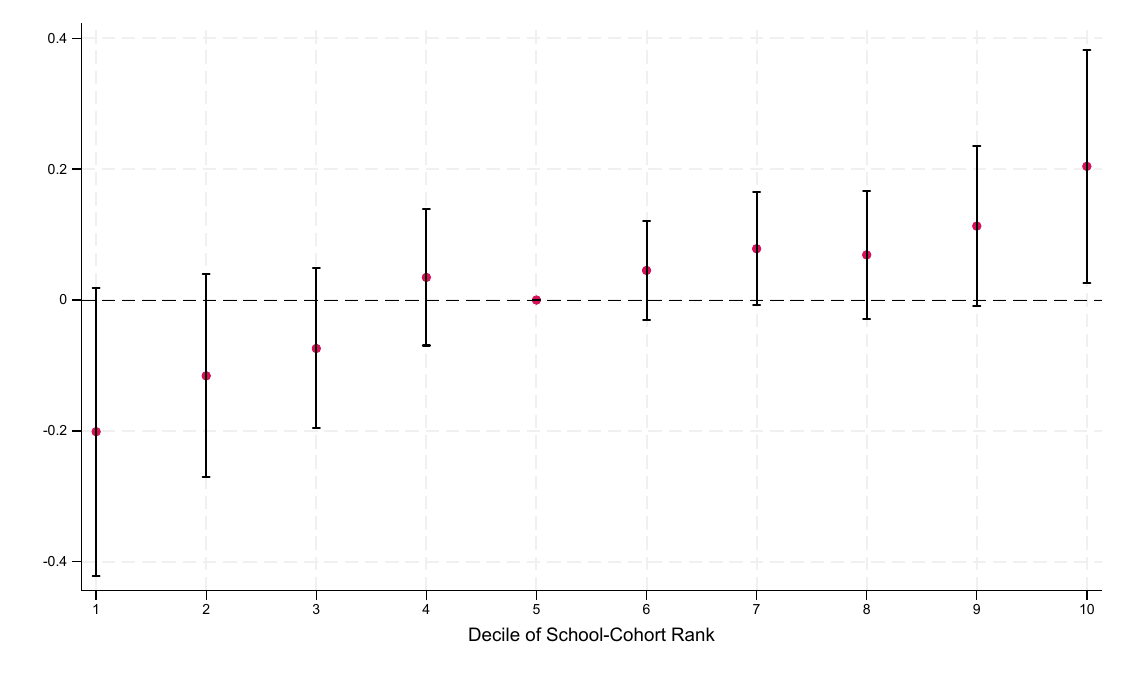}
\includegraphics[scale = 0.54]{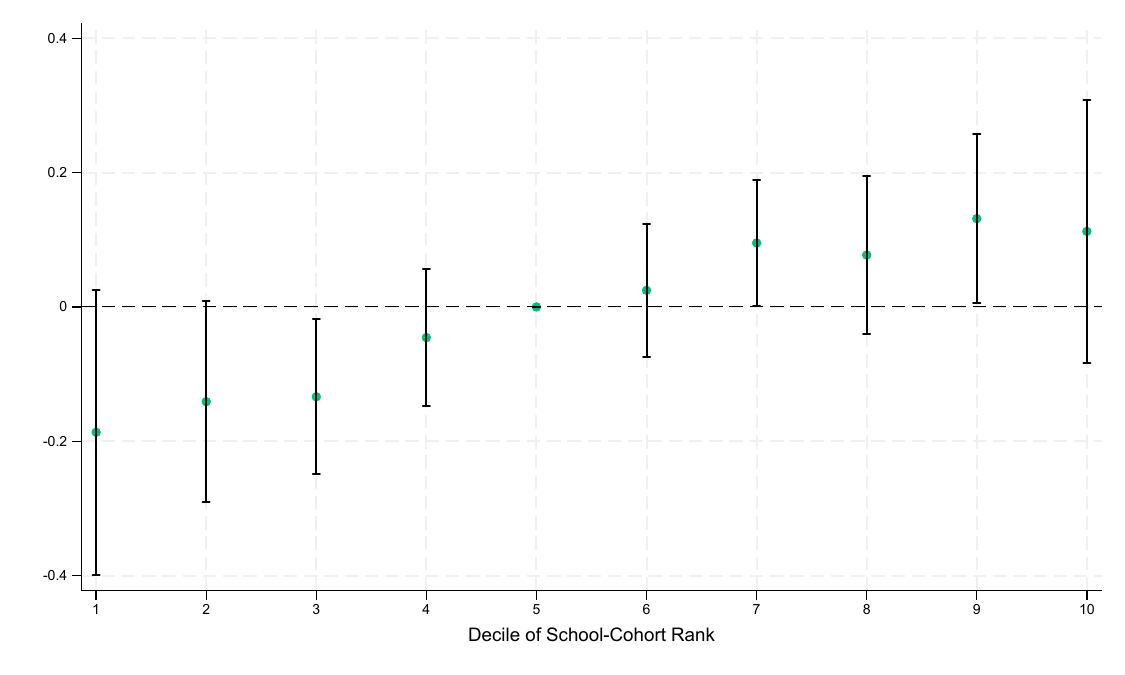}

\begin{minipage}{13 cm} \vspace{1.2ex}
\scriptsize{Notes: The graph shows the rank effect on externalizing and internalizing skills by rank decile.}
\end{minipage}

\end{figure}

\vspace{1mm}
To conclude, \autoref{graph_5} illustrates that the relationship between academic rank and both externalizing and internalizing skills follows an approximately linear trajectory, mirroring the pattern observed for academic performance. This supports the validity of our linear modeling approach and suggests that the impact of relative position on non-cognitive development is broadly proportional across the rank distribution. However, our robustness checks make clear that not all effects are equally stable: while the estimated effect on internalizing skills holds across different specifications and sample restrictions, the effect on externalizing skills appears far more fragile and dependent on the presence of a small number of outliers. Taken together, our findings suggest that rank has a broad but uneven influence on non-cognitive outcomes, with more persistent effects on students’ emotional adjustment than on their behavioral regulation.

%%%%%%%%%%%%%%%%%%%%%%%%%%%%%%%%%
% Results: Non-Cognitive Skills %
%%%%%%%%%%%%%%%%%%%%%%%%%%%%%%%%%

\setlength\extrarowheight{3pt}

\begin{table}[p]
\caption{Rank Effect on the Externalizing and Internalizing Skills} \label{table_4A}
\begin{center}
\scalebox{0.7}{
\begin{tabular}{l c c c c c c}
    
    \vspace{-0.35 cm} \\
    \hline
    \vspace{-0.35 cm} \\

                      & (1) & (2) & (3) & (4) & (5) & (6) \\
    Outcome Variables & \multicolumn{3}{c}{Externalizing Skills} & \multicolumn{3}{c}{Internalizing Skills} \\
        
    \vspace{-0.35 cm} \\
    \hline
    \vspace{-0.35 cm} \\
        
    Percentile Rank & 0.285*   & 0.272*   & 0.276*   & 0.470*** & 0.483*** & 0.487*** \\
                    & (0.157)  & (0.155)  & (0.154)  & (0.153)  & (0.153)  & (0.158)  \\
     
    \\

    Mean of the Outcome & 0     & 0     & 0     & 0     & 0     & 0     \\
    SD of the Outcome   & 1     & 1     & 1     & 1     & 1     & 1     \\
    Observations        & 6,631 & 6,516 & 6,516 & 6,631 & 6,516 & 6,516 \\
        
    \vspace{-0.35 cm} \\
    \hline
    \vspace{-0.35 cm} \\

    School Fixed Effects & X & X & X & X & X & X \\
    Cohort Fixed Effects & X & X & X & X & X & X \\
    
    \\
    
    Cognitive Skills         & X & X & X & X & X & X \\
    Cognitive Skills Squared & X & X & X & X & X & X \\

    \\

    Mean of Peer Cognitive Skills & X & X & X & X & X & X \\
    SD of Peer Cognitive Skills   & X & X & X & X & X & X \\
    
    \\
    
    Sex                  & - & X & X & - & X & X \\
    Socioeconomic Status & - & X & X & - & X & X \\
    Number of Siblings   & - & X & X & - & X & X \\
    Month of Birth       & - & X & X & - & X & X \\

    \\

    Height               & - & - & X & - & - & X \\
    Height Squared       & - & - & X & - & - & X \\
    Birth Weight         & - & - & X & - & - & X \\
    Birth Weight Squared & - & - & X & - & - & X \\
    
    \vspace{-0.35 cm} \\
    \hline
    \vspace{-0.35 cm} \\

\end{tabular}
}
\begin{minipage}{13 cm} \vspace{1.2ex}
\scriptsize{Notes: We estimate the relationship percentile rank within the school-cohort group and the standardized measures of externalizing and internalizing skills. Percentile rank is established using our baseline measure of cognitive skills, the outcome of the Age 9 Test. The individual measures of externalizing and internalizing skills are estimated using common factor analysis on the 26 items of the Rutter Questionnaire for Teachers, completed in March 1964. We include only children who were still in primary school as of March 1964; therefore, those in the 7 youngest cohorts, born between April 1952 and October 1955. We control for: a categorical variable taking a different value for each school in the sample; a categorical variable taking a different value for each cohort in the sample; a quadratic polynomial of child cognitive skills (based on the outcome of the Age 9 Test); the mean and standard deviation of the cognitive skills of the peers of the students (based on the outcome of the Age 9 Test); a quadratic polynomial of the standardized height measured during the first medical exam in school and a quadratic polynomial of birth weight; a categorical variable taking value 1 if the child is a girl, and 0 if he is a boy; a categorical variable taking value 1 if the child belongs to a family of high socioeconomic status (defined based on the father's occupation), and 0 otherwise; a categorical variable capturing the specific month of birth of the child; and the number of siblings of the child. Standard errors are clustered at the school-cohort level. *** p $<$ 0.01, ** p $<$ 0.05, * p $<$ 0.1.}
\end{minipage}
\end{center}
\end{table}

\begin{table}[p]
\caption{Removing Extreme Children with Extreme Behavioral Issues} \label{table_4B}
\begin{center}
\scalebox{0.7}{
\begin{tabular}{l c c c c c c}
    
    \vspace{-0.35 cm} \\
    \hline
    \vspace{-0.35 cm} \\

                      & (1) & (2) & (3) & (4) & (5) & (6) \\
    Outcome Variables & \multicolumn{3}{c}{Externalizing Skills} & \multicolumn{3}{c}{Internalizing Skills} \\
    No Bottom         & 0\% & 3\% & 5\% & 0\% & 3\% & 5\% \\
        
    \vspace{-0.35 cm} \\
    \hline
    \vspace{-0.35 cm} \\
        
    Percentile Rank & 0.276*   & 0.084    & 0.051    & 0.487*** & 0.335*** & 0.256*** \\
                    & (0.153)  & (0.109)  & (0.096)  & (0.158)  & (0.113)  & (0.094)  \\
     
    \\

    Mean of the Outcome & 0 & 0.13 & 0.18 & 0 & 0.11 & 0.17 \\
    SD of the Outcome   & 1 & 0.67 & 0.56 & 1 & 0.76 & 0.68 \\
    Observations        & 6,516 & 6,324 & 6,200 & 6,516 & 6,313 & 6,179 \\
        
    \vspace{-0.35 cm} \\
    \hline
    \vspace{-0.35 cm} \\

    School Fixed Effects & X & X & X & X & X & X \\
    Cohort Fixed Effects & X & X & X & X & X & X \\
    
    \\
    
    Cognitive Skills         & X & X & X & X & X & X \\
    Cognitive Skills Squared & X & X & X & X & X & X \\

    \\

    Mean of Peer Cognitive Skills & X & X & X & X & X & X \\
    SD of Peer Cognitive Skills   & X & X & X & X & X & X \\
    
    \\
    
    Sex                  & X & X & X & X & X & X \\
    Socioeconomic Status & X & X & X & X & X & X \\
    Number of Siblings   & X & X & X & X & X & X \\
    Month of Birth       & X & X & X & X & X & X \\

    \\

    Height               & X & X & X & X & X & X \\
    Height Squared       & X & X & X & X & X & X \\
    Birth Weight         & X & X & X & X & X & X \\
    Birth Weight Squared & X & X & X & X & X & X \\
    
    \vspace{-0.35 cm} \\
    \hline
    \vspace{-0.35 cm} \\

\end{tabular}
}
\begin{minipage}{13 cm} \vspace{1.2ex}
\scriptsize{Notes: We estimate the relationship percentile rank within the school-cohort group and the standardized measures of externalizing and internalizing skills, progressively excluding the children with the most severe externalizing or internalizing issues. We start from the full sample (columns (1) and (4)), then exclude the bottom 3\% (columns (2) and (5)), and finally the bottom 5\% (columns (3) and (6)). Percentile rank is established using our baseline measure of cognitive skills, the outcome of the Age 9 Test. We include only children who were still in primary school as of March 1964; therefore, those in the 7 youngest cohorts, born between April 1952 and October 1955. We control for: a categorical variable taking a different value for each school in the sample; a categorical variable taking a different value for each cohort in the sample; a quadratic polynomial of child cognitive skills (based on the outcome of the Age 9 Test); the mean and standard deviation of the cognitive skills of the peers of the students (based on the outcome of the Age 9 Test); a quadratic polynomial of the standardized height measured during the first medical exam in school and a quadratic polynomial of birth weight; a categorical variable taking value 1 if the child is a girl, and 0 if he is a boy; a categorical variable taking value 1 if the child belongs to a family of high socioeconomic status (defined based on the father's occupation), and 0 otherwise; a categorical variable capturing the specific month of birth of the child; and the number of siblings of the child. Standard errors are clustered at the school-cohort level. *** p $<$ 0.01, ** p $<$ 0.05, * p $<$ 0.1.}
\end{minipage}
\end{center}
\end{table} 

\subsection{The Rank Effect on Parental Investment} \label{sec_5.3}

\vspace{1mm}
An important hypothesis in the rank effect literature concerns the role of parents in shaping children's outcomes. In principle, a child who stands out academically might elicit greater attention and investment from parents, potentially reinforcing the effects of relative standing. However, direct evidence on the role of parental involvement in amplifying or moderating rank effects remains limited. \textcite{murphy2020} find that academic rank affects children's self-confidence but report limited evidence that it significantly alters parental behavior. Similarly, \textcite{megalokonomou2024} document that while classroom rank influences students’ self-perception and aspirations, they find little systematic impact on parental involvement. Understanding whether and how parents respond to their child's relative academic performance remains an open question.

\vspace{1mm}
To explore this dimension, we use information from the Family Survey, which involved interviewing the mothers of a randomly selected subset of children—approximately one-quarter of the full sample. Among those selected, about 80\% of families agreed to participate. As shown in \autoref{sec_8.0}, the randomization into the Family Survey was successful: children whose families participated in the survey do not differ systematically in observable characteristics from those who did not. Furthermore, when we estimate the effect of academic rank on the probability of participating in the Family Survey, we find no evidence of selection based on rank.

\vspace{1mm}
We examine the effect of academic rank on several measures of parental involvement. Specifically, we study the probability that a parent knows the child's teacher by name and expresses satisfaction with the child's academic progress; the type of support provided with homework (ranging from autonomy to active parental help); and the amount of time the child spends on homework, using dummies indicating whether the child spends less than 30/45/60 minutes a day.\footnote{Parental knowledge and satisfaction are based on whether the parent reports knowing the child's teacher's name and being happy with the child's school progress. Homework help is captured through three variables: (i) whether the child does homework autonomously, (ii) whether the parent checks or supervises the homework, and (iii) whether the parent provides active help. Parents categorize their child's homework habits among the following: ``No homework given'', ``Child does homework on their own'', ``Parent just sees it is done'', ``Parent checks homework'', or ``Parent gives active help''. We define children as autonomous as those in the ``Child does homework on their own'' category. We defined children being checked as those in the ``Parent just sees it is done'' and ``Parent checks homework'' categories. We define children receiving help as those in the ``Parent gives active help'' category. Time spent on homework is reported in intervals (e.g., 0–15 minutes, 15–30 minutes, etc.), and we create dummies indicating whether the child spends less than 30, 45, or 60 minutes per day on homework.}

\vspace{1mm}
The results reveal a very limited relationship between academic rank and parental involvement. As shown in \autoref{table_5}, there is a positive effect of rank on the probability that a parent knows the teacher’s name and is satisfied with the child's progress, significant at the 10\% level. A four-position improvement in academic rank — approximately a 10\% increase within the school-cohort group — is associated with a 2 percentage point increase in the probability of parental awareness, corresponding to about 6\% of the baseline probability. However, none of the other estimated effects on parental behavior — whether providing help with homework or influencing time spent on homework — are statistically significant.

\vspace{1mm}
Although insignificant, the signs of the estimated coefficients hint at an interesting pattern. Higher-ranked students appear slightly more likely to do homework autonomously or under minimal supervision, while lower-ranked students are somewhat more likely to receive active parental help. This could suggest a weak substitution mechanism, where parents become more directly involved when children appear to struggle. However, given the lack of statistical significance, we cannot confidently assert that parental investment systematically responds to academic rank. Moreover, it is difficult to disentangle whether differences in time spent on homework reflect changes in parental behavior or greater intrinsic motivation and commitment on the part of the child. Overall, our findings suggest that the estimated impact of rank on academic outcomes are unlikely to be primarily mediated through shifts in parental involvement.

%%%%%%%%%%%%%%%%%%%%%%%%%%%%%%%%
% Results: Parental Investment %
%%%%%%%%%%%%%%%%%%%%%%%%%%%%%%%%

\setlength\extrarowheight{3pt}

\begin{table}[p]
\caption{Rank Effect on Parental Investment} \label{table_5}
\begin{center}
\scalebox{0.7}{
\begin{tabular}{l c c c c c c c c}
    
    \vspace{-0.35 cm} \\
    \hline
    \vspace{-0.35 cm} \\

                      & (1) & (2) & (3) & (4) & (5) & (6) & (7) & (8) \\
                      & Awareness of   & Happy & \multicolumn{3}{c}{Homework:} & \multicolumn{3}{c}{Daily Time on Homework:} \\
    Outcome Variables & Teacher's Name & with Progress & Autonomous & Checked & Helped & $<$ 30 & $<$ 45 & $<$ 60 \\
        
    \vspace{-0.35 cm} \\
    \hline
    \vspace{-0.35 cm} \\
        
    Percentile Rank & 0.197* & 0.083   & 0.075   & 0.110   & -0.158  & 0.022   & -0.070  & -0.043  \\
                    & (0.117) & (0.104) & (0.111) & (0.136) & (0.126) & (0.090) & (0.131) & (0.116) \\
     
    \\

    Mean of the Outcome & 0.68  & 0.80  & 0.22 & 0.42  & 0.36  & 0.12  & 0.47  & 0.68   \\
    SD of the Outcome   & 0.47  & 0.48  & 0.41 & 0.49  & 0.50  & 0.33  & 0.50  & 0.47   \\
    Observations        & 1,787 & 1,770 & 1,787 & 1,787 & 1,787 & 1,787 & 1,787 & 1,787 \\
        
    \vspace{-0.35 cm} \\
    \hline
    \vspace{-0.35 cm} \\

    School Fixed Effects          & X & X & X & X & X & X & X & X \\
    Cohort Fixed Effects          & X & X & X & X & X & X & X & X \\
    
    \\
    
    Cognitive Skills              & X & X & X & X & X & X & X & X \\
    Cognitive Skills Squared      & X & X & X & X & X & X & X & X \\

    \\

    Mean of Peer Cognitive Skills & X & X & X & X & X & X & X & X \\
    SD of Peer Cognitive Skills   & X & X & X & X & X & X & X & X \\
    
    \\
    
    Sex                            & X & X & X & X & X & X & X & X \\
    Socioeconomic Status           & X & X & X & X & X & X & X & X \\
    Number of Siblings             & X & X & X & X & X & X & X & X \\
    Month of Birth                 & X & X & X & X & X & X & X & X \\
    
    \vspace{-0.35 cm} \\
    \hline
    \vspace{-0.35 cm} \\

\end{tabular}
}
\begin{minipage}{13 cm} \vspace{1.2ex}
\scriptsize{Notes: We estimate the relationship percentile rank within the school-cohort group and different outcomes. Percentile rank is established using our baseline measure of cognitive skills, the outcome of the Age 9 Test. These outcomes are derived from the interviews conducted in the context of the Family Survey. The different outcomes are: the probability that the parents know the teacher's name (column (1)); the probability that the parents are happy about the child's progress in school (column (2)); the probability that the child does his homework autonomously (column (3)), with the parent checking them (column (4)), or with the parent's help (column (5)); the probability that the parents reports that the child spends a daily time on homework below 30 minutes (column (6)), below 45 minutes (column (7)), or below 60 minutes (column (8)). Our sample consists of 10 cohorts of children who attended primary school in Aberdeen in December 1962, who were born between October 1950 and October 1955, and who participated in the Family Survey. We control for: a categorical variable taking a different value for each school in the sample; a categorical variable taking a different value for each cohort in the sample; a quadratic polynomial of child cognitive skills (based on the outcome of the Age 9 Test); the mean and standard deviation of the cognitive skills of the peers of the students (based on the outcome of the Age 9 Test); a categorical variable taking value 1 if the child is a girl, and 0 if he is a boy; a categorical variable taking value 1 if the child belongs to a family of high socioeconomic status (defined based on the father's occupation), and 0 otherwise; a categorical variable capturing the specific month of birth of the child; and the number of siblings of the child. Standard errors are clustered at the school-cohort level.}
\end{minipage}
\end{center}
\end{table}

\subsection{The Rank Effect on Long-Term Outcomes} \label{sec_5.4}

\vspace{1mm}
We next examine the long-term consequences of academic rank, using responses from a follow-up survey conducted by mail in 2001, nearly 40 years after the original data collection. The response rate for the follow-up was approximately 60\%. While the existing literature has documented short- and medium-term effects of rank on academic and non-cognitive outcomes, evidence on long-term impacts remains limited. \textcite{denning2023} find persistent effects on earnings 25 years after school, but in general, the literature tends to assume that early impacts on education and non-cognitive skills naturally translate into longer-run differences without direct empirical confirmation.

\vspace{1mm}
Comparing observable characteristics, we show in \autoref{sec_8.0} that respondents to the 2001 follow-up were more likely to be women, to come from higher socioeconomic backgrounds, to have had better grades and stronger externalizing skills in primary school, and to come from smaller families. However, crucially, there are no significant differences in academic rank between respondents and non-respondents. While sample selection could bias levels of long-term outcomes, the fact that it is unrelated to rank supports the validity of our estimates for studying the causal impact of academic rank. Moreover, we control for a wide range of baseline characteristics in all regressions, further mitigating concerns about differential selection.

\vspace{1mm}
We start by analyzing how individuals retrospectively recall their primary school experience.\footnote{Primary school experience is captured through three variables. First, self-reported happiness in primary school, based on the question ``Were you happy in primary school?'' with possible answers being ``Very happy''. ``Fairly happy'', ``Neither happy or unhappy'', ``Not very happy'', and ``Not at all happy''. We create a dummy equal to one for respondents reporting being either ``Very happy'' or ``Fairly happy''. Second, perceptions of social relationships are assessed through the question ``How many friends did you have in primary school compared to other children?'', allowing for the answers ``More'', ``Same number'', or ``Fewer''. We construct two dummies for having more or fewer friends. Third, to investigate exposure to bullying, we use a dummy for affirmative responses to ``Were you bullied in primary school?''.} We also study educational attainment.\footnote{Educational achievement is measured using several binary indicators: (i) attendance at grammar school (a selective secondary school for higher-achieving students based on the 11-plus exam results), (ii) attainment of O-level qualifications (exams typically taken at age 16), (iii) attainment of A-level qualifications (advanced exams taken at age 18), and (iv) attainment of a university degree.}

\vspace{1mm}
The results, presented in \autoref{table_6A}, show mixed evidence. Rank has a positive and statistically significant effect on the probability of recalling being happy in primary school, significant at the 5\% level. A four-position improvement in rank—about a 10\% shift in the group—raises the likelihood of reporting happiness by approximately 1.5\% relative to the baseline. However, we find no significant effects of rank on the recalled number of friends or on reported experiences of being bullied.

\vspace{1mm}
As expected, the effects of academic rank are much more pronounced on educational outcomes. A four-position increase in rank is associated with a 4 percentage point rise in the probability of attending grammar school — an increase of roughly 20\% relative to the baseline rate. Similarly, we find positive and significant effects on achieving O-level and A-level qualifications, with a four-position jump corresponding to a 5\% and 12\% increase in probability, respectively. In contrast, we find no evidence that academic rank affects the likelihood of completing a university degree: the estimated coefficient is close to zero and imprecisely measured. This may reflect historical context, as university attendance was considerably less common during the period when these cohorts reached adulthood, and admission standards may have been stricter or less influenced by early academic differences.

\vspace{1mm}
We next examine the impact of academic rank on socioeconomic status, earnings, family formation, and subjective well-being in adulthood.\footnote{Socioeconomic status takes value one if the occupation of an individual is ``Legislators, senior officials, and managers'', and zero otherwise. It is based on the one-digit occupational code from ISCO-88. Income is reported in brackets, with thresholds at £1,000, £4,000, £8,000, £12,500, £17,500, £25,000, £35,000, and £45,000. We examine the probability of earning more than £25,000, £35,000, and £45,000 per year.} We also consider the probability of having ever married or having children, based on self-reported answers to direct questions about marital and parental status.\footnote{Participants answered ``Have you ever been married?'' and ``Do you have kids?'', which we use to define dummy variables representing the probability of marriage and of having kids.} Finally, we include two measures of subjective well-being: enjoyment of daily activities and general happiness.\footnote{Participants were asked ``How do you feel you enjoy daily activities?'' and ``Do you feel happy in your everyday life?'' with possible responses being ``More than usual'', ``Same as usual'', ``Less than usual'', and ``Much less than usual''. We construct two dummies: one indicating whether the participant enjoys daily activities the same or more than usual, and another for feeling happy the same or more than usual.}

\vspace{1mm}
Results are reported in \autoref{table_6B}. Overall, we find limited evidence that academic rank has a meaningful long-term effect on socioeconomic status or income. While the estimated coefficients are generally positive, they are imprecisely measured and generally statistically insignificant. We detect a modest effect of rank on the probability of earning more than £25,000 per year, significant at the 10\% level: a four-position improvement in school-cohort rank corresponds to a 2.2\% increase in the likelihood of earning above this threshold relative to the baseline probability. In 2001, the median household income in the UK was approximately £20,000 \parencite{ONS2003}, making the £25,000 threshold a reasonable proxy for identifying individuals in the upper half of the income distribution. However, given the lack of significant effects at higher income brackets (£35,000 and £45,000) and the wide confidence intervals around the estimates, this result does not provide strong evidence of a robust long-term earnings effect.

\vspace{1mm}
Turning to family outcomes, we find no significant effect of academic rank on the probability of marrying or having children. Similarly, there is no evidence that higher academic rank affects adult subjective well-being, as measured by enjoyment of daily activities or general happiness. In each case, the point estimates are small and statistically insignificant, suggesting that any potential impact of primary school rank on these aspects of adult life is minimal or absent.

\vspace{1mm}
Overall, our analysis indicates that while academic rank has persistent effects on educational achievement, its influence on broader long-term outcomes appears to be limited. It is important to acknowledge that the sample used for the 2001 follow-up survey consists largely of individuals who remained in Aberdeen or the surrounding areas, which may attenuate the estimated effects on socioeconomic and life outcomes. The implications of this selection are not entirely clear: if higher-achieving individuals were more likely to migrate out of Aberdeen and subsequently less likely to respond to the follow-up, we may be missing a segment of the population for whom the rank effect could be particularly impactful. On the other hand, negative selection into the follow-up sample — where respondents are relatively less mobile or lower achieving — could therefore contribute to the limited long-term effects we observe.

%%%%%%%%%%%%%%%%%%%%%%%%%%%%%%%
% Results: Long-term Outcomes %
%%%%%%%%%%%%%%%%%%%%%%%%%%%%%%%

\setlength\extrarowheight{3pt}

\begin{table}[p]
\caption{Rank Effect on Long-Term Outcomes: Primary School Memories and Academic Achievement} \label{table_6A}
\begin{center}
\scalebox{0.7}{
\begin{tabular}{l c c c c c c c c}
    
    \vspace{-0.35 cm} \\
    \hline
    \vspace{-0.35 cm} \\

                      & (1) & (2) & (3) & (4) & (5) & (6) & (7) & (8) \\
                      & \multicolumn{4}{c}{Primary School Memories:} & \multicolumn{4}{c}{Academic Achievement:} \\
    Outcome Variables & Happy & More Friends & Less Friends & Bullied & Grammar School & O-Level & A-Level & Degree \\
        
    \vspace{-0.35 cm} \\
    \hline
    \vspace{-0.35 cm} \\
        
    Percentile Rank & 0.128** & 0.024   & -0.037  & 0.011   & 0.398*** & 0.295*** & 0.411*** & 0.004   \\
                    & (0.057) & (0.033) & (0.041) & (0.069) & (0.054)  & (0.065)  & (0.058)  & (0.047) \\
     
    \\

    Mean of the Outcome & 0.81  & 0.05  & 0.09  & 0.24  & 0.20  & 0.60  & 0.33  & 0.15  \\
    SD of the Outcome   & 0.39  & 0.21  & 0.28  & 0.43  & 0.40  & 0.49  & 0.47  & 0.36  \\
    Observations        & 6,631 & 6,516 & 6,516 & 6,631 & 6,516 & 6,516 & 6,516 & 6,516 \\
        
    \vspace{-0.35 cm} \\
    \hline
    \vspace{-0.35 cm} \\

    School Fixed Effects          & X & X & X & X & X & X & X & X \\
    Cohort Fixed Effects          & X & X & X & X & X & X & X & X \\
    
    \\
    
    Cognitive Skills              & X & X & X & X & X & X & X & X \\
    Cognitive Skills Squared      & X & X & X & X & X & X & X & X \\

    \\

    Mean of Peer Cognitive Skills & X & X & X & X & X & X & X & X \\
    SD of Peer Cognitive Skills   & X & X & X & X & X & X & X & X \\
    
    \\
    
    Sex                            & X & X & X & X & X & X & X & X \\
    Socioeconomic Status           & X & X & X & X & X & X & X & X \\
    Number of Siblings             & X & X & X & X & X & X & X & X \\
    Month of Birth                 & X & X & X & X & X & X & X & X \\
    
    \vspace{-0.35 cm} \\
    \hline
    \vspace{-0.35 cm} \\

\end{tabular}
}
\begin{minipage}{13 cm} \vspace{1.2ex}
\scriptsize{Notes: We estimate the relationship percentile rank within the school-cohort group and different outcomes. Percentile rank is established using our baseline measure of cognitive skills, the outcome of the Age 9 Test. These outcomes are derived from the answers to the 2001 Follow-Up survey. The different outcomes are: the probability that the participant recalls being happy during primary school (column (1)); the probability that the participant recalls having more friends than the other children during primary school (column (2)); the probability that the participant recalls having less friends than the other children during primary school (column (3)); the probability that the participant recalls being bullied during primary school (column (4)); the probability that the participant reports having attended a grammar school (column (5)); the probability that the participant reports having achieved O-level education (column (6)); the probability that the participant reports having achieved A-level education (column (7)); the probability that the participant reports having achieved a degree (column (8)). Our sample consists of 10 cohorts of children who attended primary school in Aberdeen in December 1962, who were born between October 1950 and October 1955, and participated in the 2001 mail Follow-Up to the original survey. We control for: a categorical variable taking a different value for each school in the sample; a categorical variable taking a different value for each cohort in the sample; a quadratic polynomial of child cognitive skills (based on the outcome of the Age 9 Test); the mean and standard deviation of the cognitive skills of the peers of the students (based on the outcome of the Age 9 Test); a categorical variable taking value 1 if the child is a girl, and 0 if he is a boy; a categorical variable taking value 1 if the child belongs to a family of high socioeconomic status (defined based on the father's occupation), and 0 otherwise; a categorical variable capturing the specific month of birth of the child; and the number of siblings of the child. Standard errors are clustered at the school-cohort level.}
\end{minipage}
\end{center}
\end{table}

\begin{table}[p]
\caption{Rank Effect on Long-Term Outcomes: Socioeconomic Status, Earnings, Fertility, and Well-Being} \label{table_6B}
\begin{center}
\scalebox{0.7}{
\begin{tabular}{l c c c c c c c c}
    
    \vspace{-0.35 cm} \\
    \hline
    \vspace{-0.35 cm} \\

                      & (1) & (2) & (3) & (4) & (5) & (6) & (7) & (8) \\
                      &     & \multicolumn{3}{c}{Annual Income Above:} & \multicolumn{2}{c}{Probability of Having:} & \multicolumn{2}{c}{Well-Being:} \\
    Outcome Variables & SES & £25,000 & £35,000 & £45,000 & Married & Children & Enjoy Day & Happy \\
        
    \vspace{-0.35 cm} \\
    \hline
    \vspace{-0.35 cm} \\
        
    Percentile Rank & 0.048   & 0.118*  & 0.034   & 0.028   & 0.018   & -0.026  & 0.052   & 0.071   \\
                    & (0.052) & (0.065) & (0.062) & (0.047) & (0.043) & (0.554) & (0.414) & (0.050) \\
     
    \\

    Mean of the Outcome & 0.17  & 0.43  & 0.24  & 0.11  & 0.92  & 0.86  & 0.89  & 0.88  \\
    SD of the Outcome   & 0.37  & 0.50  & 0.43  & 0.32  & 0.27  & 0.35  & 0.31  & 0.33  \\
    Observations        & 6,631 & 6,516 & 6,516 & 6,631 & 6,516 & 6,516 & 6,516 & 6,516 \\
        
    \vspace{-0.35 cm} \\
    \hline
    \vspace{-0.35 cm} \\

    School Fixed Effects          & X & X & X & X & X & X & X & X \\
    Cohort Fixed Effects          & X & X & X & X & X & X & X & X \\
    
    \\
    
    Cognitive Skills              & X & X & X & X & X & X & X & X \\
    Cognitive Skills Squared      & X & X & X & X & X & X & X & X \\

    \\

    Mean of Peer Cognitive Skills & X & X & X & X & X & X & X & X \\
    SD of Peer Cognitive Skills   & X & X & X & X & X & X & X & X \\
    
    \\
    
    Sex                            & X & X & X & X & X & X & X & X \\
    Socioeconomic Status           & X & X & X & X & X & X & X & X \\
    Number of Siblings             & X & X & X & X & X & X & X & X \\
    Month of Birth                 & X & X & X & X & X & X & X & X \\
    
    \vspace{-0.35 cm} \\
    \hline
    \vspace{-0.35 cm} \\

\end{tabular}
}
\begin{minipage}{13 cm} \vspace{1.2ex}
\scriptsize{Notes: We estimate the relationship percentile rank within the school-cohort group and different outcomes. Percentile rank is established using our baseline measure of cognitive skills, the outcome of the Age 9 Test. These outcomes are derived from the answers to the 2001 Follow-Up survey, to which roughly 60\% of the participants in the original survey responded. The different outcomes are: the probability of being in a high socioeconomic status, which is defined based on the occupation code (column (1)); the probability that the participant reports an annual income above £25,000 (column (2)), above £35,000 (column (3)), or above £45,000 (column (4)); the probability of the participant reports having ever married (column (5)); the probability of the participant reports having ever had children (column (6)); the probability of the participant reports enjoying daily activities (column (7)); the probability of the participant reports being happy (column (8)). Our sample consists of 10 cohorts of children who attended primary school in Aberdeen in December 1962, who were born between October 1950 and October 1955, and participated in the 2001 mail Follow-Up to the original survey. We control for: a categorical variable taking a different value for each school in the sample; a categorical variable taking a different value for each cohort in the sample; a quadratic polynomial of child cognitive skills (based on the outcome of the Age 9 Test); the mean and standard deviation of the cognitive skills of the peers of the students (based on the outcome of the Age 9 Test); a categorical variable taking value 1 if the child is a girl, and 0 if he is a boy; a categorical variable taking value 1 if the child belongs to a family of high socioeconomic status (defined based on the father's occupation), and 0 otherwise; a categorical variable capturing the specific month of birth of the child; and the number of siblings of the child. Standard errors are clustered at the school-cohort level.}
\end{minipage}
\end{center}
\end{table}

\subsection{Gender Heterogeneity} \label{sec_5.5}

\vspace{1mm}
Gender is a natural dimension along which to explore heterogeneity in the rank effect, given the pronounced differences in boys’ and girls’ cognitive and non-cognitive profiles during childhood. In our sample, girls exhibit stronger cognitive skills on average (mean of 0.09 vs -0.08 for boys), significantly higher externalizing skills (0.18 vs -0.17), and slightly lower internalizing skills (-0.05 vs 0.05). These initial disparities may shape how boys and girls respond to their relative academic standing within their school-cohort groups.

\vspace{1mm}
\autoref{table_7A} presents gender-specific estimates of the rank effect. We find that the impact on academic performance is approximately 30\% larger for girls than for boys. For non-cognitive outcomes, boys drive the rank effect on externalizing skills, beginning from a lower baseline and exhibiting a stronger response to higher relative rank. In contrast, girls show a substantially larger and more precisely estimated rank effect on internalizing skills, consistent with their lower initial levels of these traits. However, we show in \autoref{table_7B} that the effect on externalizing skills is highly sensitive to the exclusion of extreme cases: when we remove the bottom 3–5\% of students with the most severe externalizing behaviors, the estimated impact becomes small and statistically insignificant for both boys and girls. By contrast, the rank effect on internalizing skills remains robust across these sample restrictions, retaining statistical significance and meaningful effect sizes for both genders. This provides further evidence that the overall rank effect is primarily driven by changes in self-concept and emotional adjustment, as captured by internalizing skills, rather than by improvements in behavioral regulation.

\vspace{1mm}
These gender differences extend to other outcomes as well. In terms of parental investment, the small but statistically significant overall effect on parental awareness of the teacher’s name appears to be driven by girls, though the gender difference is not itself statistically significant. For educational attainment, rank boosts the probability of grammar school attendance for both genders, but the effect is larger for girls (22\% vs 14\% increase relative to their respective baselines for a 4-position jump within the school-cohort group). Interestingly, the rank effect on completing O-levels is about 50\% higher for boys than for girls, while the effect on A-levels is nearly identical. These findings likely reflect the gender norms of the period, where girls who excelled academically were more likely to pursue selective secondary education but faced greater constraints in progressing to university.

\vspace{1mm}
Finally, the modest positive effect of rank on adult earnings observed in the full sample appears to be concentrated among boys. For them, a four-position increase in rank leads to a 1.8 percentage point increase in the probability of earning more than £25,000 per year (2.8\% of the baseline), statistically significant at the 5\% level. This effect is absent for girls, and the gender difference is itself statistically significant. These results suggest that while academic rank has broadly positive effects for both genders, the long-term economic returns may have been more accessible to boys in this historical context, possibly due to gendered labor market structures and educational pathways. For the other outcomes explored in previous sections, gender heterogeneity does not appear to play a major role.

%%%%%%%%%%%%%%%%%%%%%%%%%%%%%%%%%
% Results: Gender Heterogeneity %
%%%%%%%%%%%%%%%%%%%%%%%%%%%%%%%%%

\setlength\extrarowheight{3pt}

\begin{table}[p]
\caption{Gender Heterogeneity} \label{table_7A}
\begin{center}
\scalebox{0.65}{
\begin{tabular}{l c c c c c c c c}
    
    \vspace{-0.35 cm} \\
    \hline
    \vspace{-0.35 cm} \\

                      & (1) & (2) & (3) & (4) & (5) & (6) & (7) & (8) \\
                      &  & \multicolumn{2}{c}{Non-Cognitive Skills:} & Awareness of & \multicolumn{3}{c}{Academic Achievement:} & Annual Income: \\
    Outcome Variables & VRT & Externalizing & Internalizing & Teacher's Name & Grammar & O-Level & A-Level & $>$ £25,000 \\
        
    \vspace{-0.35 cm} \\
    \hline
    \vspace{-0.35 cm} \\

     & \multicolumn{8}{c}{Rank Effect for Boys} \\

    \vspace{-0.35 cm} \\
    \hline
    \vspace{-0.35 cm} \\
        
    Percentile Rank & 0.497*** & 0.421** & 0.288* & 0.166  & 0.250*** & 0.343*** & 0.405*** & 0.183** \\
                    & (0.072) & (0.167) & (0.162) & (0.123) & (0.056)  & (0.066)  & (0.062)  & (0.071) \\
                    
    \\

    Mean of the Outcome & -0.04 & -0.17 & 0.05  & 0.65 & 0.18  & 0.34  & 0.21  & 0.63  \\
    SD of the Outcome   & 1.01  & 1.16  & 0.96  & 0.48 & 0.38  & 0.47  & 0.41  & 0.48  \\
    Observations        & 4,902 & 3,349 & 3,349 & 925  & 2,844 & 2,844 & 2,844 & 2,844 \\

    \vspace{-0.35 cm} \\
    \hline
    \vspace{-0.35 cm} \\

     & \multicolumn{8}{c}{Rank Effect for Girls} \\

    \vspace{-0.35 cm} \\
    \hline
    \vspace{-0.35 cm} \\
    
    Percentile Rank & 0.645*** & 0.119   & 0.603*** & 0.220* & 0.492*** & 0.232*** & 0.413*** & 0.074   \\
                    & (0.069)  & (0.152) & (0.160)  & (0.119) & (0.056)  & (0.068)  & (0.058)  & (0.065) \\

    \\

    Mean of the Outcome & 0.04   & 0.18   & -0.05  & 0.71   & 0.22   & 0.37   & 0.18   & 0.25    \\
    SD of the Outcome   & 0.98   & 0.75   & 1.03   & 0.46   & 0.42   & 0.48   & 0.38   & 0.43    \\
    Observations        & 4,539  & 3,167  & 3,167  & 900    & 3,031  & 3,031  & 3,031  & 3,031   \\

    \vspace{-0.35 cm} \\
    \hline
    \vspace{-0.35 cm} \\
    
    T-test of the Difference & -0.148*** & 0.303***  & -0.315*** & -0.054   & -0.242*** & 0.111*** & -0.008  & 0.109** \\
                             & (0.035)   & (0.083)   & (0.091)   & (0.063) & (0.029)   & (0.038)  & (0.034) & (0.043) \\
        
    \vspace{-0.35 cm} \\
    \hline
    \vspace{-0.35 cm} \\

    School Fixed Effects          & X & X & X & X & X & X & X & X \\
    Cohort Fixed Effects          & X & X & X & X & X & X & X & X \\
    
    \\
    
    Cognitive Skills              & X & X & X & X & X & X & X & X \\
    Cognitive Skills Squared      & X & X & X & X & X & X & X & X \\

    \\

    Mean of Peer Cognitive Skills & X & X & X & X & X & X & X & X \\
    SD of Peer Cognitive Skills   & X & X & X & X & X & X & X & X \\
    
    \\
    
    Sex                            & X & X & X & X & X & X & X & X \\
    Socioeconomic Status           & X & X & X & X & X & X & X & X \\
    Number of Siblings             & X & X & X & X & X & X & X & X \\
    Month of Birth                 & X & X & X & X & X & X & X & X \\

    \\

    Height                         & - & X & X & - & - & - & - & - \\
    Height Squared                 & - & X & X & - & - & - & - & - \\
    Birth Weight                   & - & X & X & - & - & - & - & - \\
    Birth Weight Squared           & - & X & X & - & - & - & - & - \\
    
    \vspace{-0.35 cm} \\
    \hline
    \vspace{-0.35 cm} \\

\end{tabular}
}
\begin{minipage}{13 cm} \vspace{1.2ex}
\scriptsize{Notes: We estimate the relationship percentile rank within the school-cohort group and different outcomes, providing the linear combination of the effect of rank based and of the interaction between the rank and the children's gender. Percentile rank is established using our baseline measure of cognitive skills, the outcome of the Age 9 Test. The sample size changes depending on the outcome estimated, as they are derived from different surveys. The outcomes are: the standardized score of the Verbal Reasoning Test, or ``VRT''  (column (1)); the standardized measures of externalizing and internalizing skills (columns (2) and (3)); the probability that the parents know the teacher's name (column (4)); the probability that the participant reports having attended a grammar school (column (5)); the probability that the participant reports having achieved O-level (columns (6)) or A-level education (column (7)); the probability that the participant reports an annual income above £25,000 (column (8)). We control for: a categorical variable taking a different value for each school in the sample; a categorical variable taking a different value for each cohort in the sample; a quadratic polynomial of child cognitive skills (based on the outcome of the Age 9 Test); the mean and standard deviation of the cognitive skills of the peers of the students (based on the outcome of the Age 9 Test); a quadratic polynomial of the standardized height measured during the first medical exam in school and a quadratic polynomial of birth weight; a categorical variable taking value 1 if the child is a girl, and 0 if he is a boy; we control for a categorical variable taking value 1 if the child belongs to a family of high socioeconomic status (defined based on the father's occupation), and 0 otherwise; we control for a categorical variable capturing the specific month of birth of the child; finally, we control for the number of siblings of the child. Standard errors are clustered at the school-cohort level. *** p $<$ 0.01, ** p $<$ 0.05, * p $<$ 0.1.}
\end{minipage}
\end{center}
\end{table}

\begin{table}[p]
\caption{Gender Heterogeneity: Removing Extreme Children with Extreme Behavioral Issues} \label{table_7B}
\begin{center}
\scalebox{0.7}{
\begin{tabular}{l c c c c c c}
    
    \vspace{-0.35 cm} \\
    \hline
    \vspace{-0.35 cm} \\

                      & (1) & (2) & (3) & (4) & (5) & (6) \\
    Outcome Variables & \multicolumn{3}{c}{Externalizing Skills} & \multicolumn{3}{c}{Internalizing Skills} \\
    No Bottom         & 0\% & 3\% & 5\% & 0\% & 3\% & 5\% \\
        
    \vspace{-0.35 cm} \\
    \hline
    \vspace{-0.35 cm} \\

     & \multicolumn{6}{c}{Rank Effect for Boys} \\

    \vspace{-0.35 cm} \\
    \hline
    \vspace{-0.35 cm} \\
        
    Percentile Rank & 0.421**  & 0.160    & 0.105    & 0.288*   & 0.234*   & 0.165*  \\
                    & (0.167)  & (0.110)  & (0.099)  & (0.162)  & (0.119)  & (0.098) \\
     
    \\

    Mean of the Outcome & -0.17 & 0.03  & 0.10  & 0.05  & 0.15  & 0.20 \\
    SD of the Outcome   & 1.16  & 0.75  & 0.62  & 0.96  & 0.75  & 0.66 \\
    Observations        & 3,349 & 3,275 & 3,183 & 3,349 & 3,329 & 3,261 \\

    \vspace{-0.35 cm} \\
    \hline
    \vspace{-0.35 cm} \\

     & \multicolumn{6}{c}{Rank Effect for Girls} \\

    \vspace{-0.35 cm} \\
    \hline
    \vspace{-0.35 cm} \\
        
    Percentile Rank & 0.119    & 0.021    & 0.006    & 0.603*** & 0.423*** & 0.333*** \\
                    & (0.152)  & (0.113)  & (0.098)  & (0.160)  & (0.115)  & (0.097)  \\
     
    \\

    Mean of the Outcome & 0.18  & 0.24  & 0.27  & -0.05 & 0.08  & 0.13  \\
    SD of the Outcome   & 0.75  & 0.56  & 0.49  & 1.03  & 0.77  & 0.69  \\
    Observations        & 3,167 & 3,187 & 3,151 & 3,167 & 3,125 & 3,057 \\

    \vspace{-0.35 cm} \\
    \hline
    \vspace{-0.35 cm} \\
    
    School Fixed Effects & X & X & X & X & X & X \\
    Cohort Fixed Effects & X & X & X & X & X & X \\
    
    \\
    
    Cognitive Skills         & X & X & X & X & X & X \\
    Cognitive Skills Squared & X & X & X & X & X & X \\

    \\

    Mean of Peer Cognitive Skills & X & X & X & X & X & X \\
    SD of Peer Cognitive Skills   & X & X & X & X & X & X \\
    
    \\
    
    Sex                  & X & X & X & X & X & X \\
    Socioeconomic Status & X & X & X & X & X & X \\
    Number of Siblings   & X & X & X & X & X & X \\
    Month of Birth       & X & X & X & X & X & X \\

    \\

    Height               & X & X & X & X & X & X \\
    Height Squared       & X & X & X & X & X & X \\
    Birth Weight         & X & X & X & X & X & X \\
    Birth Weight Squared & X & X & X & X & X & X \\
    
    \vspace{-0.35 cm} \\
    \hline
    \vspace{-0.35 cm} \\

\end{tabular}
}
\begin{minipage}{13 cm} \vspace{1.2ex}
\scriptsize{Notes: We estimate the relationship percentile rank within the school-cohort group and the standardized measures of externalizing and internalizing skills, providing the linear combination of the effect of rank based and of the interaction between the rank and the children's gender, and progressively excluding the children with the most severe externalizing or internalizing issues. We start from the full sample (columns (1) and (4)), then exclude the bottom 3\% (columns (2) and (5)), and finally the bottom 5\% (columns (3) and (6)). We control for: a categorical variable taking a different value for each school in the sample; a categorical variable taking a different value for each cohort in the sample; a quadratic polynomial of child cognitive skills (based on the outcome of the Age 9 Test); the mean and standard deviation of the cognitive skills of the peers of the students (based on the outcome of the Age 9 Test); a quadratic polynomial of the standardized height measured during the first medical exam in school and a quadratic polynomial of birth weight; a categorical variable taking value 1 if the child is a girl, and 0 if he is a boy; we control for a categorical variable taking value 1 if the child belongs to a family of high socioeconomic status (defined based on the father's occupation), and 0 otherwise; we control for a categorical variable capturing the specific month of birth of the child; finally, we control for the number of siblings of the child. Standard errors are clustered at the school-cohort level. *** p $<$ 0.01, ** p $<$ 0.05, * p $<$ 0.1.}
\end{minipage}
\end{center}
\end{table}

\subsection{Robustness of the Results: Using an Alternative Identifying Variation} \label{sec_5.6}

\vspace{1mm}
The identifying variation we exploit comes from differences in academic rank between school-cohort groups, leveraging the quasi-random assignment of students to these groups. Our baseline approach controls for school and cohort fixed effects separately, effectively assuming that any unobserved characteristics specific to a school (e.g., teaching quality, peer norms) or cohort (e.g., year-specific shocks, changes in policy) can be controlled for additively. While our balancing tests and randomization checks provide strong evidence that this assumption holds in our setting, it remains a relatively strong restriction, as it implies that no unobserved factor simultaneously influences both the school and cohort components in a way that is correlated with students’ academic rank.

\vspace{1mm}
To assess the robustness of our findings to this assumption, we re-estimate our main results using school-cohort fixed effects, which absorb all common variation within each school-cohort cell. This approach captures any shared characteristics that might jointly affect all students within a given peer group, controlling for group-specific unobserved heterogeneity. While this specification sacrifices some statistical power by narrowing the scope of identifying variation to differences between rather than within school-cohort groups, it provides a more conservative but stringent test of the rank effect.

\vspace{1mm}
Main results from the specification using school-cohort fixed effects are presented in \autoref{table_R1}. Overall, this alternative approach yields slightly larger point estimates for most outcomes, though the differences are small and not statistically distinguishable from those obtained with separate school and cohort fixed effects. The only notable exception is the coefficient on income, which becomes smaller and statistically insignificant when school-cohort fixed effects are used. However, the gender-specific analysis reveals that the previously identified rank effect on boys’ earnings remains robust — both in magnitude and statistical significance — even under this more stringent specification. Taken together, the results indicate that the choice of fixed effects structure has only a minor impact on our estimated effects, lending further credibility to our main findings.

%%%%%%%%%%%%%%%%%%%%%%%
% Robustness Exercise %
%%%%%%%%%%%%%%%%%%%%%%%

\setlength\extrarowheight{3pt}

\begin{table}[p]
\caption{Robustness Exercise: Using School-Cohort Fixed Effects} \label{table_R1}
\begin{center}
\scalebox{0.65}{
\begin{tabular}{l c c c c c c c c}
    
    \vspace{-0.35 cm} \\
    \hline
    \vspace{-0.35 cm} \\

                      & (1) & (2) & (3) & (4) & (5) & (6) & (7) & (8) \\
    Outcome Variables & VRT & Externalizing & Internalizing & Teacher's Name & Grammar & O-Level & A-Level & Income $>$ £25,000 \\
        
    \vspace{-0.35 cm} \\
    \hline
    \vspace{-0.35 cm} \\
        
    Percentile Rank & 0.612*** & 0.287*  & 0.525*** & 0.228* & 0.432*** & 0.322*** & 0.417*** & 0.092 \\
                    & (0.068)  & (0.158) & (0.158)  & (0.132) & (0.055)  & (0.069)  & (0.060)  & (0.068) \\
                    
    \\

    Mean of the Outcome & 0 & 0 & 0 & 0.68 & 0.20 & 0.60 & 0.33 & 0.43 \\
    SD of the Outcome   & 1 & 1 & 1 & 0.47 & 0.40 & 0.49 & 0.47 & 0.50 \\
    Observations        & 9,441 & 6,516 & 6,516 & 1,787 & 5,744 & 5,744 & 5,744 & 5,744 \\
        
    \vspace{-0.35 cm} \\
    \hline
    \vspace{-0.35 cm} \\

    School-Cohort Fixed Effects   & X & X & X & X & X & X & X & X \\
    
    \\
    
    Cognitive Skills              & X & X & X & X & X & X & X & X \\
    Cognitive Skills Squared      & X & X & X & X & X & X & X & X \\

    \\

    Mean of Peer Cognitive Skills & X & X & X & X & X & X & X & X \\
    SD of Peer Cognitive Skills   & X & X & X & X & X & X & X & X \\
    
    \\
    
    Sex                            & X & X & X & X & X & X & X & X \\
    Socioeconomic Status           & X & X & X & X & X & X & X & X \\
    Number of Siblings             & X & X & X & X & X & X & X & X \\
    Month of Birth                 & X & X & X & X & X & X & X & X \\

    \\

    Height                         & - & X & X & - & - & - & - & - \\
    Height Squared                 & - & X & X & - & - & - & - & - \\
    Birth Weight                   & - & X & X & - & - & - & - & - \\
    Birth Weight Squared           & - & X & X & - & - & - & - & - \\
    
    \vspace{-0.35 cm} \\
    \hline
    \vspace{-0.35 cm} \\

\end{tabular}
}
\begin{minipage}{13 cm} \vspace{1.2ex}
\scriptsize{Notes: We estimate the relationship percentile rank within the school-cohort group and different outcomes. Percentile rank is established using our baseline measure of cognitive skills, the outcome of the Age 9 Test. The sample size changes depending on the outcome estimated, as they are derived from different surveys. The outcomes are: the standardized score of the Verbal Reasoning Test, or ``VRT''  (column (1)); the standardized measures of externalizing and internalizing skills (columns (2) and (3)); the probability that the parents know the teacher's name (column (4)); the probability that the participant reports having attended a grammar school (column (5)); the probability that the participant reports having achieved O-level (columns (6)) or A-level education (column (7)); the probability that the participant reports an annual income above £25,000 (column (8)). We control for: a categorical variable taking a different value for each school-cohort group in the sample; a quadratic polynomial of child cognitive skills (based on the outcome of the Age 9 Test); the mean and standard deviation of the cognitive skills of the peers of the students (based on the outcome of the Age 9 Test); a quadratic polynomial of the standardized height measured during the first medical exam in school and a quadratic polynomial of birth weight; a categorical variable taking value 1 if the child is a girl, and 0 if he is a boy; a categorical variable taking value 1 if the child belongs to a family of high socioeconomic status (defined based on the father's occupation), and 0 otherwise; a categorical variable capturing the specific month of birth of the child; and the number of siblings of the child. Standard errors are clustered at the school-cohort level.}
\end{minipage}
\end{center}
\end{table}

%%%%%%%%%%%%%%
% Conclusion %
%%%%%%%%%%%%%%

\section{Conclusion} \label{sec_6.0}

\vspace{1mm}
This paper examines the consequences of academic rank within school-cohort peer groups, using rich data on the entire population of children enrolled in Aberdeen, Scotland, primary schools in 1962. Exploiting quasi-random variation in peer group composition, we identify the causal impact of a student’s relative academic standing, conditional on ability, on a wide range of short- and long-term outcomes. We leverage detailed school records, teacher-reported behavioral assessments, randomized family interviews, and a follow-up survey conducted nearly four decades later to provide a comprehensive view of how rank shapes trajectories from childhood to adulthood.

\vspace{1mm}
We find that academic rank has a substantial effect on students’ academic performance. A 10\% improvement in relative position — equivalent to a four-place jump in rank within the average group — raises test scores on the high-stakes 11-plus exam by around 6\% of a standard deviation. These effects are larger for girls, who also begin with higher baseline academic achievement. We also show that the same increase in rank improves internalizing skills, such as confidence and self-concept, by around 4.5\% of a standard deviation. This effect is robust across specifications and not driven by extreme behavioral cases. In contrast, the impact on externalizing skills (e.g., impulse control, attention) is weaker and sensitive to sample restrictions. These findings suggest that rank shapes how students perceive themselves more than how they regulate their behavior.

\vspace{1mm}
Parental investment does not appear to systematically adjust in response to rank. While higher-ranked children are slightly more likely to have parents who know their teacher’s name, we detect no meaningful changes in homework support or time allocation. This suggests that rank effects operate primarily through children’s internal responses rather than shifts in parental behavior. In the long term, we find that academic rank continues to shape educational attainment, especially for girls. Rank increases the probability of attending grammar school and completing formal secondary qualifications (O-levels and A-levels). However, these educational gains do not translate uniformly into labor market outcomes. Rank improves the probability of having an income above £25,000 only for boys, and we detect no effect on occupational status, marriage, fertility, or adult well-being. The gender asymmetry in long-term outcomes likely reflects the historical context, when educational opportunities for women were expanding, but labor market access remained restricted.

\vspace{1mm}
These results make several contributions to the literature. We provide the most detailed evidence to date on the role of academic rank in shaping non-cognitive development. Unlike previous studies that focus on single-trait, self-reported outcomes such as self-confidence or expectations \parencite[e.g.,][]{murphy2020, elsner2021}, we use validated teacher-reported measures and apply factor analysis to capture broader constructs of internalizing and externalizing behavior. Our findings highlight that rank has a particularly robust effect on internalizing traits, aligning with psychological theories that link social comparison to identity formation and self-efficacy.

\vspace{1mm}
We also find empirical evidence on the long-term consequences of academic rank, nearly 40 years after the reference point. While recent work has established effects on educational choice and earnings in the short run \parencite[e.g.,][]{denning2023, goulas2022}, we show that these effects persist well into adulthood — but unevenly so. Educational returns are more visible for girls, while income effects are concentrated among boys. These patterns point to important constraints — social, institutional, and historical — on the extent to which academic advantages can translate into broader life outcomes. More broadly, our findings reinforce the idea that ranking within a peer group is not just a reflection of performance but a determinant of future development, with implications for how we structure peer interactions in schools.

%%%%%%%%%%%%%%%%
% Bibliography %
%%%%%%%%%%%%%%%%

\newpage
\printbibliography

% Flush main body tables and figures %
\processdelayedfloats

%%%%%%%%%%%%%%
% Appendices %
%%%%%%%%%%%%%%

\newpage
\appendix

% Reset table and figure counters %

\setcounter{table}{0}
\renewcommand{\thetable}{A\arabic{table}}
\setcounter{figure}{0}
\renewcommand{\thefigure}{A\arabic{figure}}

\newpage

%%%%%%%%%%%%%%
% Appendix A %
%%%%%%%%%%%%%%

\section*{Appendix A: Randomization of Participation to the Family Survey and 2001 Follow-Up} \label{sec_8.0}

\vspace{1mm}
In this appendix, we present additional checks to support the validity of our identification strategy for the estimation of the rank effect on parental investment and long-term outcomes. First, we show that academic rank does not predict the probability of participating in the Family Survey, and that observable child characteristics are balanced between participants and non-participants, confirming the effectiveness of the original randomization. We then replicate this exercise for the 2001 follow-up survey, demonstrating that rank is not associated with survey participation. However, individual baseline characteristics are unbalanced across respondents and non-respondents.

\vspace{1mm}
In \autoref{table_A1}, we test whether academic rank predicts participation in the Family Survey or the 2001 follow-up survey. Across both outcomes, we find no evidence that rank influences the probability of participation. This result is reassuring, as it suggests that our estimates of the rank effect on parental investment and long-term outcomes are not biased by differential attrition related to students’ relative standing within their school-cohort group. In other words, even though participation in these surveys is selective, that selection appears orthogonal to the variation we exploit for identification.

\vspace{1mm}
To further assess the nature of selection into these survey subsamples, we regress a range of baseline individual characteristics on the probability of participating in either the Family Survey or the 2001 follow-up, as reported in \autoref{table_A2}. These characteristics include cognitive skills (measured by the age 9 test), non-cognitive skills (externalizing and internalizing scores), and other observables such as gender, socioeconomic background, height and weight residuals from the first medical exam, birth weight, and number of siblings.\footnote{Specifically: a dummy for the child being female, a high-SES indicator (based on the parent's occupation), residualized height and weight from the first medical exam, birth weight in pounds, and the number of siblings as of December 1962.} The results show that participation in the Family Survey is effectively randomized: the only statistically significant coefficient is for number of siblings, and even that is marginal (10\% level) and economically small - 1\% of a unit of standard deviation. 

\vspace{1mm}
In contrast, participation in the 2001 follow-up survey is clearly non-randomized. Respondents tend to have stronger cognitive and non-cognitive skills, are more likely to be female, and come from higher-SES families. There are also differences in anthropometric measures and family size, though the latter remains of limited economic relevance. These patterns suggest that the long-term sample is positively selected, with higher-performing children more likely to respond. While this does not undermine the internal validity of our estimates — since rank itself does not predict participation — it implies that the external validity of our long-term results may be limited to a more advantaged segment of the original sample. We limit these concerns by controlling for these individual characteristics in our main specifications.

%%%%%%%%%%%%%%
% Appendix A %
%%%%%%%%%%%%%%

\setlength\extrarowheight{3pt}

\begin{table}[p]
\caption{Rank Effect on Family Survey and 2001 Follow-Up Survey Participation} \label{table_A1}
\begin{center}
\scalebox{0.7}{
\begin{tabular}{l c c c c}
    
    \vspace{-0.35 cm} \\
    \hline
    \vspace{-0.35 cm} \\

                      & (1) & (2) & (3) & (4) \\
                      & \multicolumn{4}{c}{Participation to the:} \\
    Outcome Variables & \multicolumn{2}{c}{Family Survey} & \multicolumn{2}{c}{2001 Follow-Up} \\
        
    \vspace{-0.35 cm} \\
    \hline
    \vspace{-0.35 cm} \\
        
    Percentile Rank & 0.012   & 0.013   & 0.046   & 0.045   \\
                    & (0.044) & (0.044) & (0.056) & (0.056) \\
    \\

    Mean of the Outcome & 0.18  & 0.18  & 0.59  & 0.59  \\
    SD of the Outcome   & 0.39  & 0.39  & 0.49  & 0.49  \\
    Observations        & 9,715 & 9,715 & 9,715 & 9,715 \\
        
    \vspace{-0.35 cm} \\
    \hline
    \vspace{-0.35 cm} \\

    School Fixed Effects & X & X & X & X \\
    Cohort Fixed Effects & X & X & X & X \\
    
    \\
    
    Cognitive Skills         & X & X & X & X \\
    Cognitive Skills Squared & X & X & X & X \\

    \\

    Mean of Peer Cognitive Skills & X & X & X & X \\
    SD of Peer Cognitive Skills   & X & X & X & X \\
    
    \\
    
    Sex                  & - & X & - & X \\
    Socioeconomic Status & - & X & - & X \\
    Number of Siblings   & - & X & - & X \\
    Month of Birth       & - & X & - & X \\
    
    \vspace{-0.35 cm} \\
    \hline
    \vspace{-0.35 cm} \\

\end{tabular}}

\begin{minipage}{13 cm} \vspace{1.2ex}
\scriptsize{Notes: We estimate the relationship percentile rank within the school-cohort group and the probability of participating to the Family Survey (columns (1) and (2)) and to the 2001 Follow-Up Survey (columns (3) and (4)). Percentile rank is established using our baseline measure of cognitive skills, the outcome of the Age 9 Test. Our sample consists of 10 cohorts of children who attended primary school in Aberdeen in December 1962, who were born between October 1950 and October 1955. We control for: a categorical variable taking a different value for each school in the sample; a categorical variable taking a different value for each cohort in the sample; a quadratic polynomial of child cognitive skills (based on the outcome of the Age 9 Test); the mean and standard deviation of the cognitive skills of the peers of the students (based on the outcome of the Age 9 Test); a categorical variable taking value 1 if the child is a girl, and 0 if he is a boy; a categorical variable taking value 1 if the child belongs to a family of high socioeconomic status (defined based on the father's occupation), and 0 otherwise; a categorical variable capturing the specific month of birth of the child; and the number of siblings of the child. Standard errors are clustered at the school-cohort level.}
\end{minipage}

\end{center}
\end{table}

\begin{table}[p]
\caption{Balancing Exercise: Probability of Participating in the Family Survey and the 2001 Follow-Up Survey} \label{table_A2}
\begin{center}
\scalebox{0.65}{

\begin{tabular}{l c c c c c c c c c} 

    \vspace{-0.35 cm} \\
    \hline
    \vspace{-0.35 cm} \\
    
    Independent Variables & Cognitive Skills & Externalizing & Internalizing & Woman & High SES & Height & Weight & Birth Weight & Siblings \\

    \vspace{-0.35 cm} \\
    \hline
    \vspace{-0.35 cm} \\

     \multicolumn{10}{c}{Relationship between individual characteristics and the probability of participating in the Family Survey} \\
     
    \vspace{-0.35 cm} \\
    \hline
    \vspace{-0.35 cm} \\
     
    Individual Characteristics & 0.001   & 0.007   & -0.003  & 0.010   & -0.016  & 0.005   & 0.002   & 0.003   & -0.010* \\
                               & (0.004) & (0.004) & (0.005) & (0.008) & (0.014) & (0.004) & (0.004) & (0.004) & (0.005) \\

    \vspace{-0.35 cm} \\
    \hline
    \vspace{-0.35 cm} \\

     \multicolumn{10}{c}{Relationship between individual characteristics and the probability of participating in the 2001 Follow-Up Survey} \\ 
     
    \vspace{-0.35 cm} \\
    \hline
    \vspace{-0.35 cm} \\
     
    Individual Characteristics & 0.085*** & 0.060*** & 0.007*** & 0.087*** & -0.069*** & 0.038*** & 0.025*** & 0.005   & -0.019*** \\
                               & (0.005)  & (0.006)  & (0.006)  & (0.010)  & (0.017)   & (0.005)  & (0.005)  & (0.005) & (0.007)   \\
     
    \vspace{-0.35 cm} \\
    \hline
    \vspace{-0.35 cm} \\

    Observations & 9,717 & 6,790 & 6,790 & 9,970 & 9,970 & 9,666 & 9,668 & 9,970 & 9,970 \\ 

    \vspace{-0.35 cm} \\
    \hline
    \vspace{-0.35 cm} \\

\end{tabular}}

\begin{minipage}{13 cm} \vspace{1.2ex}
\scriptsize{Notes: We estimate the unconditional relationship between different individual characteristics and the probability of participating in the Family Survey and the 2001 Follow-Up Survey. These characteristics are: the student cognitive skills, measured by the standardized score in the Age 9 Test; the standardized measures of externalizing and internalizing skills, estimated using common factor analysis on the 26 items of the Rutter Questionnaire for Teachers, completed in March 1964; the student probability of being a girl, the student probability of coming from an advantaged socioeconomic background (based on the father's occupation), the student height and weight at the time of their first medical exam, the student birth weight (lbs), and the student number of siblings. We include all 10 cohorts of children who attended primary school in Aberdeen in December 1962, who were born between October 1950 and October 1955. Standard errors are clustered at the school-cohort-group level. *** p $<$ 0.01, ** p $<$ 0.05, * p $<$ 0.1.}
\end{minipage}

\end{center}
\end{table}

% Flush main body tables and figures %
\processdelayedfloats

\end{document}